\documentclass[usenatbib]{mnras}
\usepackage{natbib}
\usepackage{amssymb}
\usepackage{aas_macros}
\usepackage{url}
\usepackage{amsmath}
\usepackage{fixltx2e}
\usepackage{amsmath}

\newcommand{\Msol}{\ensuremath{\text{M}_{\odot}}}

\newcommand{\rt}{\ensuremath{R_{\text{200}}}}
\newcommand{\rf}{\ensuremath{R_{\text{500}}}}

\newcommand{\mf}{\ensuremath{M_{\text{500}}}}

\newcommand{\etal}{et al.\ }

\newcommand{\Chandra}{\emph{Chandra}}

\newcommand{\XMM}{\emph{XMM-Newton}}

\newcommand{\gta}{\,\rlap{\raise 0.4ex\hbox{$>$}}{\lower 0.6ex\hbox{$\sim$}}\,}
\newcommand{\lta}{\,\rlap{\raise 0.4ex\hbox{$<$}}{\lower 0.6ex\hbox{$\sim$}}\,}

\newcommand{\km}{\mbox{\ensuremath{\text{~km}}}}

\newcommand{\Mpc}{\mbox{\ensuremath{\text{~Mpc}}}}
\newcommand{\s}{\mbox{\ensuremath{\text{~s}}}}

\newcommand{\Gyr}{\mbox{\ensuremath{\text{~Gyr}}}}

\newcommand{\pMpc}{\ensuremath{{\Mpc^{-1}}}}
\newcommand{\ps}{\ensuremath{\s^{-1}}}

\newcommand{\kmpspMpc}{\ensuremath{{\km \ps \pMpc\,}}}

\newcommand{\apropto}{\mathrel{\vcenter{
  \offinterlineskip\halign{\hfil$##$\cr
    \propto\cr\noalign{\kern2pt}\sim\cr\noalign{\kern-2pt}}}}}
\newcommand{\dex}{\mbox{\ensuremath{\text{~dex}}}}

\usepackage[utf8]{inputenc}
\usepackage[T1]{fontenc}
\usepackage{fixltx2e}
\usepackage{graphicx}
\usepackage{grffile}
\usepackage{longtable}
\usepackage{wrapfig}
\usepackage{rotating}
\usepackage[normalem]{ulem}
\usepackage{amsmath}
\usepackage{textcomp}
\usepackage{amssymb}
\usepackage{capt-of}
\usepackage{hyperref}
\usepackage{enumitem}
\setitemize{noitemsep,topsep=5pt,parsep=5pt,partopsep=0pt,leftmargin=10pt}
\usepackage{setspace}
\usepackage{listings}
\usepackage{amsmath}

\newcommand{\MX}{\ensuremath{M_{\text{X}}}}
\newcommand{\MC}{\ensuremath{M_{\text{C}}}}
\newcommand{\MWL}{\ensuremath{M_{\text{WL}}}}

\newcommand{\sigx}{\ensuremath{\sigma_{\text{X}}}}
\newcommand{\sigc}{\ensuremath{\sigma_{\text{C}}}}
\newcommand{\delx}{\ensuremath{\delta_{\text{X}}}}
\newcommand{\delc}{\ensuremath{\delta_{\text{C}}}}
\newcommand{\kapx}{\ensuremath{\kappa_{\text{X}}}}
\newcommand{\kapc}{\ensuremath{\kappa_{\text{C}}}}
\newcommand{\muX}{\ensuremath{\mu_{\text{X}}}}
\newcommand{\muC}{\ensuremath{\mu_{\text{C}}}}
\newcommand{\muXh}{\ensuremath{\hat{\mu}_{\text{X}}}}
\newcommand{\muCh}{\ensuremath{\hat{\mu}_{\text{C}}}}

\usepackage[dvipsnames]{xcolor}

\author{Ben Maughan}
\date{\today}
\title{}
\hypersetup{
 pdfauthor={Ben Maughan},
 pdftitle={},
 pdfkeywords={},
 pdfsubject={},
 pdfcreator={Emacs 24.5.1 (Org mode 8.3.4)},
 pdflang={English}}
\begin{document}

\title[Hydrostatic and caustic mass profiles]{Hydrostatic and Caustic Mass Profiles of Galaxy Clusters.}
\author[Ben J. Maughan \etal]
  {Ben J. Maughan,$^{1}$\thanks{E-mail: ben.maughan@bristol.ac.uk}
    Paul A. Giles,$^1$ Kenneth J. Rines,$^{2,3}$ Antonaldo Diaferio,$^{4,5}$
    \and Margaret J. Geller$^{3}$, Nina Van Der Pyl$^1$ and Massimiliano Bonamente$^{6,7}$ \medskip \\
  $^1$H. H. Wills Physics Laboratory, University of Bristol, Tyndall Ave, Bristol BS8 1TL, UK\\
  $^2$Department of Physics \& Astronomy, Western Washington University, Bellingham, WA 98225, USA\\
  $^3$Smithsonian Astrophysical Observatory, 60 Garden St, MS 20, Cambridge, MA 02138, USA\\
  $^4$Dipartimento di Fisica, Universit\`a di Torino, Via P. Giuria 1, I-10125 Torino, Italy\\
  $^5$Istituto Nazionale di Fisica Nucleare (INFN), sezione di Torino, Via P. Giuria 1, I-10125 Torino, Italy\\
  $^6$Department of Physics, University of Alabama in Huntsville, Huntsville, AL 35899, USA\\
  $^7$NASA National Space Science and Technology Center, Huntsville, AL 35812, USA
}

%\date{}
\maketitle

\begin{abstract}
We compare X-ray and caustic mass profiles for a sample of 16 massive galaxy clusters. We assume hydrostatic equilibrium in interpreting the X-ray data, and use large samples of cluster members with redshifts as a basis for applying the caustic technique. The hydrostatic and caustic masses agree to better than \(\approx20\%\) on average across the radial range covered by both techniques \((\sim[0.2-1.25]\rf)\). The mass profiles were measured independently and do not assume a common functional form. Previous studies suggest that, at \(\rf\), the hydrostatic and caustic masses are biased low and high respectively. We find that the ratio of hydrostatic to caustic mass at \(\rf\) is \(1.20^{+0.13}_{-0.11}\); thus it is larger than 0.9 at \(\approx3\sigma\) and the combination of under- and over-estimation of the mass by these two techniques is \(\approx10\%\) at most. There is no indication of any dependence of the mass ratio on the X-ray morphology of the clusters, indicating that the hydrostatic masses are not strongly systematically affected by the dynamical state of the clusters. Overall, our results favour a small value of the so-called hydrostatic bias due to non-thermal pressure sources.
\end{abstract}


\section{Introduction}
\label{sec:orgheadline1}
The observational determination of the masses of galaxy clusters is of
central importance to our understanding of the growth of structure in
the Universe and the use of clusters as cosmological probes.
Furthermore, cluster mass is an essential reference point for studies
of the astrophysical processes shaping the properties of the baryons
in clusters, both the intra-cluster medium (ICM) and the member
galaxies.

The task of measuring cluster masses is challenging, as their dominant
dark matter component can only be studied indirectly. The total mass
of a given cluster can be determined either by measuring the effect of
its gravitational potential on the properties of its ICM and galaxies,
or its gravitational lensing effect on the light from background
sources.

The most accurate and precise mass estimation techniques include
hydrostatic masses determined from X-ray observations of the ICM
\citep{sar86,mar98b,dav01,vik06a} caustic techniques based on galaxy
dynamics \citep{dia97,rin03,rin13,gif13a}, and weak gravitational lensing
measurements \citep{tys90,mel99,oka10a,hoe15}. These methods require
spatially-resolved measurements with high data quality (large numbers
of X-ray photons, galaxy redshifts, or lensed sources are needed).
Less direct mass proxies include X-ray luminosity or temperature
\citep[e.g.][]{rei02,mau07b,man10a,boh14}, and cluster richness
\citep[e.g.][]{roz08a,and10,sza11,ryk14}. These lower quality mass proxies
are calibrated against the more reliable measurements.

Historically, X-ray hydrostatic masses have been the gold standard for
calibrating other techniques, but departures from hydrostatic
equilibrium or the presence of non-thermal pressure sources (such as
turbulence, bulk motions of the ICM or cosmic rays) can lead to biases
in the estimated mass. Hydrodynamical simulations suggest that
hydrostatic masses underestimate the true mass by \(10-30\%\)
\citep{ras06,nag07,lau09,ras12,nel14a}. Observational evidence for
departures from hydrostatic equilibrium has been seen for the outer
parts of A1835, where the inferred hydrostatic cumulative mass profile
starts to decrease un-physically with radius \citep{bon13}. In
addition, uncertainties in the absolute calibration of \(\XMM\) and
\(\Chandra\) could result in biased temperature estimates leading to
biased hydrostatic mass estimates \citep[e.g.][]{mah13,roz14,sch15}.
However, we note that \citet{mar14} found excellent agreement between
hydrostatic masses derived from \(\Chandra\) and \(\XMM\) for clusters
with data from both observatories.

Recently, the question of biases in hydrostatic mass estimates has
received a great deal of attention as more sophisticated approaches
and improved data have significantly reduced the systematic
uncertainties on weak lensing masses. Several recent studies have
compared weak-lensing and hydrostatic masses (sometimes indirectly
through Sunyaev-Zel'dovich effect scaling relations calibrated with
hydrostatic masses), finding a wide range of estimates for the amount
of bias in hydrostatic masses. For example \citet{von14},
\citet{don14}, \citet{ser15b} and \citet{hoe15} found hydrostatic
masses to be biased low by \(\sim 20-30\%\), while \citet{gru14}, \citet{isr14},
\citet{app16} and \citet{smi16} found no significant evidence for
biases in the hydrostatic masses relative to weak lensing masses
\citep[with the possible exception of clusters at $z>0.3$;][but see also Israel et al. (2014)]{smi16}. The underestimation of hydrostatic
masses could account for some of the tension between the cosmological
constraints from the \emph{Planck} cosmic microwave background and cluster
number counts experiments \citep{pla14f,pla14d}. At least some of the
variation in estimates of the hydrostatic bias can be explained by
differences in redshift range and analysis techniques used
\citep[see][]{smi16}, but it remains unclear at present if there is a
significant bias in hydrostatic mass estimates.

Mass profile estimates from applying the caustic technique to galaxy
redshift data provide an attractive alternative to weak gravitational
lensing as a means of investigating biases in hydrostatic masses. The
caustic method identifies the characteristic structure in the
line-of-sight velocity and projected-radius space that traces the
escape velocity profile of a cluster, and hence can be used to
reconstruct the enclosed mass to radii well beyond the virial radius
\citep{dia97,dia99,ser11,gif13a}. Like lensing measurements, caustic
masses are independent of the dynamical state of the cluster, and are
insensitive to the physical processes that might cause the hydrostatic
biases. Caustic masses are subject to a completely different set of
systematic uncertainties than lensing masses and provide a useful
independent test to lensing-based studies.

Comparisons between hydrostatic and caustic mass profiles are rare,
with the only previous such study limited to three clusters
\citep{dia05}. Here we compare X-ray hydrostatic and caustic mass
profiles for 16 massive clusters spanning a range of dynamical states.
In this study, we examine the ratio of the two mass estimators as a
function of cluster radius for the full sample and for subsets of
relaxed and non-relaxed clusters.

The analysis assumes a WMAP9 cosmology \(H_0=69.3\kmpspMpc\),
\(\Omega_m=0.29\), \(\Omega_\Lambda=0.71\) \citep{hin13}.

\section{Cluster Sample}
\label{sec:orgheadline2}
We identify clusters from the Hectospec Cluster survey
\citep[HECS;][]{rin13}, that are also included in the complete
\emph{Chandra} sample of X-ray luminous clusters from \citet{lan13}. This
gives an overlap of 16 clusters, summarised in Table \ref{tab:orgtable1}. The
coordinates given in Table \ref{tab:orgtable1} are those of the original
X-ray survey data from which the HeCS clusters were selected
\citep{rin13}. All but one of the clusters came from the X-ray
flux-limited subset of the HeCS; A2631 is a lower flux cluster that
was also observed as part of the HeCS.

\begin{table*}
\caption{\label{tab:orgtable1}
The cluster sample. \(N\) indicates the number of cluster members with measured redshifts for the dynamical analysis.  The \emph{Chandra} exposure time after cleaning of the lightcurves is given, along with the \emph{Chandra} obsIDs used in the analysis. \(^\dagger\text{A2631}\) is not part of the flux-limited HeCS sample.}
\centering
\begin{tabular}{lrrrrrl}
\hline
Cluster & RA & Dec & \(z\) & \(N\) & exposure (ks) & obsID\\
\hline
A0267 & 28.1762 & 1.0125 & 0.2291 & 226 & 7 & 1448\\
A0697 & 130.7362 & 36.3625 & 0.2812 & 185 & 17 & 4217\\
A0773 & 139.4624 & 51.7248 & 0.2173 & 173 & 40 & 533,3588,5006\\
A0963 & 154.2600 & 39.0484 & 0.2041 & 211 & 36 & 903\\
A1423 & 179.3420 & 33.6320 & 0.2142 & 230 & 36 & 538,11724\\
A1682 & 196.7278 & 46.5560 & 0.2272 & 151 & 20 & 11725\\
A1763 & 203.8257 & 40.9970 & 0.2312 & 237 & 20 & 3591\\
A1835 & 210.2595 & 2.88010 & 0.2506 & 219 & 193 & 6880,6881,7370\\
A1914 & 216.5068 & 37.8271 & 0.1660 & 255 & 19 & 3593\\
A2111 & 234.9337 & 34.4156 & 0.2291 & 208 & 31 & 544,11726\\
A2219 & 250.0892 & 46.7058 & 0.2257 & 461 & 118 & 14355,14356,14431\\
A2261 & 260.6129 & 32.1338 & 0.2242 & 209 & 24 & 5007\\
A2631\(^\dagger\) & 354.4206 & 0.2760 & 0.2765 & 173 & 26 & 3248,11728\\
RXJ1720 & 260.0370 & 26.6350 & 0.1604 & 376 & 45 & 1453,3224,4361\\
RXJ2129 & 322.4186 & 0.0973 & 0.2339 & 325 & 40 & 552,9370\\
Zw3146 & 155.9117 & 4.1865 & 0.2894 & 106 & 79 & 909,9371\\
\hline
\end{tabular}
\end{table*}

\section{Analysis}
\label{sec:orgheadline6}
\subsection{X-ray data}
\label{sec:orgheadline3}
The \emph{Chandra} data analysis is described in \citep{gil15a},
which presents the X-ray scaling relations of the \citet{lan13}
sample. The analysis closely follows that of \citet{mau12}, but we
summarise the main steps here. The data were reduced and analysed with
version 4.6 of the CIAO software
package\footnote{\url{http://asc.harvard.edu/ciao/}},
using calibration
database\footnote{\url{http://cxc.harvard.edu/caldb/}}
version 4.5.9. Projected temperature profiles of the ICM were measured
from spectra extracted in annular regions centred on the X-ray
centroid. Similarly, projected emissivity profiles were measured from
the X-ray surface brightness in annular regions with the same centre.

Hydrostatic mass profiles \(\MX(R)\) were derived following the method
of \citet{vik06a}, assuming functional forms for the 3D density and
temperature profiles of the cluster gas, and then projecting these to
fit to the observed projected temperature and emissivity profiles. The
best-fitting 3D profiles were then used to compute the hydrostatic
mass profiles.

The statistical uncertainties on the hydrostatic mass profiles were
determined with a Monte-Carlo approach \citep{vik06a,gil15a}.
Synthetic data points were generated for the projected temperature and
emissivity profiles by sampling from the best fitting models (after
projection) at the radii of the original data. The samples were drawn
from Gaussian distributions centred on the model value with a standard
deviation given by the fractional measurement error on the original
data at each point. The same fractional error was used to assign the
error bar to the synthetic point.

The synthetic data were then fit in the same way as the original data,
and the process was repeated 1,000 times, yielding 1,000 synthetic
mass profiles. The uncertainty, \(\Delta(\MX)\), on the hydrostatic mass
at any radius was then computed as
\begin{align}
\frac{\Delta(\MX)}{\MX} = \frac{\text{sd}(\tilde{M}_{X})}{\langle \tilde{M}_{X}\rangle}
\end{align}
where \(\tilde{M}_X\) indicates the synthetic mass profiles, and
\(\text{sd}(\tilde{M}_{X})\) and \(\langle \tilde{M}_{X}\rangle\) are the
standard deviation and mean of the synthetic profile realisations
respectively.

As described in \citet{gil15a}, clusters were also classed as relaxed,
cool core clusters (hereafter RCC) if they had a low central cooling
time \((<7.7\Gyr)\), a peaked density profile (with a logarithmic slope
\(>0.7\) in the core), and a low centroid shift (\(<0.009\), indicating
regularity of X-ray isophotes). These criteria are defined and
justified in \citet{gil15a}, but see also e.g.
\citet{moh93,hud10,mau12} for related discussions. This definition is
fairly conservative. Only 5/16 clusters are RCC. The remaining 11 are
termed NRCC, but two of these (A0963 and A2261) fail only one of the
three criteria.

\subsection{Galaxy caustic masses}
\label{sec:orgheadline4}
HeCS is a spectroscopic survey of X-ray selected clusters with
MMT/Hectospec \citep{fab05}. HeCS uses the caustic technique to
measure mass profiles from large numbers of redshifts (\(\sim200\)
members per cluster; Table \ref{tab:orgtable1}). Galaxies in cluster infall
regions occupy overdense envelopes in phase-space diagrams of
line-of-sight velocity versus projected radius. The edges of these
envelopes trace the escape velocity profile of the cluster and can
therefore be used to determine the cluster mass profile. \citet{dia97}
show that the mass of a spherical shell within the infall region is
the integral of the square of the caustic amplitude \(A(r)\):
\begin{align}
\label{causticstomass}
GM(<R) - GM(<R_0) =  \mathcal{F}_\beta \int_{R_0}^{R} A^2(R)\,dR
\end{align}
where \(\mathcal{F}_\beta \simeq 0.5\) is a filling factor with a value
estimated from numerical simulations \citep{dia99}. We approximate
\(\mathcal{F}_\beta\) as a constant; variations in \(\mathcal{F}_\beta\)
with radius lead to some systematic uncertainty in the mass profile we
derive from the caustic technique. In particular, the caustic mass
profile assuming constant \(\mathcal{F}_\beta\) may overestimate the
true mass profile within \(\sim 0.5 \rt\) in simulated clusters by \(\sim
15\%\) or more \citep{ser11}. We include these issues in our assessment
of the intrinsic uncertainties and biases in the technique
\citep{ser11}. HeCS used the algorithm of \citet{dia99} to identify
the amplitude of the caustics and determine the cluster mass profiles.

The uncertainties on the caustic masses were derived from the
uncertainty in the caustic location \citep{dia99}. Clusters like A0697
(with large uncertainties) have an irregular phase space diagram
with a poorly defined edge. The clusters with small uncertainties
contain large numbers of members and sharply defined edges in phase
space. These errors reflect the statistical precision of
the measurement; there is expected to be a \(\sim30\%\) intrinsic
scatter between caustic mass and true mass \citep{ser11}.

\subsection{Modelling the mass biases}
\label{sec:orgheadline5}
With the mass profiles in hand, we then modelled the biases in the
hydrostatic and caustic mass profiles in terms of the ratio \(\MX/\MC\).
Note that by convention when we report masses (i.e. in Fig. \ref{fig:orgparagraph1}
and Table \ref{tab.m500}), we express them and their uncertainty as
the mean (\(M\)) and standard deviation (\(S\)) of the probability
distribution determined from the analyses in Sections \ref{sec:orgheadline3} and
\ref{sec:orgheadline4} in \emph{linear space}. This facilitates comparisons
with other work. However, when modelling the biases in the masses the
likelihood of the observed masses are assumed to be lognormal (in
base 10) with mean \(\mu\) and standard deviation \(\sigma\). These are
related to \(M\) and \(S\) by
\begin{align}
\mu & = \log_{10}\left(\frac{M}{\sqrt{1+S^2/M^2}}\right) \label{eq.mu} \\
\sigma & = \sqrt{\log_{10}\left(\frac{S^2}{M^2} + 1 \right)} \label{eq.sig}.
\end{align}
The choice of a lognormal rather than normal distribution for the
likelihood of the observed masses is motivated by the following
reasons. First, the distribution of masses in the error analysis of
the X-ray and caustic masses more closely resembles a lognormal than
normal distribution. Second, the ratio of lognormally distributed
quantities itself follows a lognormal distribution, while the ratio of
normally distributed quantities follows a Cauchy distribution, which
has undefined moments making the resulting uncertainty on \(\MX/\MC\)
harder to interpret.

In order to constrain the bias and scatter between the two mass
estimators, we performed a Bayesian analysis. We constructed a model
in which a given cluster has observed hydrostatic and caustic masses
\(\muXh\) and \(\muCh\), respectively (we use \(\mu\) throughout to
signify logarithmic masses, and the hats indicate that these are
observed quantities). These observed masses are related to the "true"
hydrostatic and caustic masses \(\muX\) and \(\muC\) by the following
stochastic relations
\begin{align}
\muXh \sim \cal{N}(\muX,\sigx) \\
\muCh \sim \cal{N}(\muC,\sigc)
\end{align}
where "\(\sim\)" means "is distributed as" and \(\sigx\) and \(\sigc\) are the
standard deviations of lognormal likelihoods  describing
the observed hydrostatic and caustic masses, respectively. \(\cal{N}\)
denotes a normal distribution. The \(\hat{\mu}\) and \(\sigma\) values are computed from the masses and errors given in Table \ref{tab.m500} using Eqs \ref{eq.mu} and \ref{eq.sig}.

These mass proxies are then related to the real mass of the cluster
\(\mu\) (again in base 10 log space) by the stochastic relations
\begin{align}
\muX \sim \cal{N}(\mu + \kapx,\delx) \\
\muC \sim \cal{N}(\mu + \kapc,\delc)
\end{align}
where \(\kapx\) and \(\kapc\) parametrise the bias between the real mass
and the hydrostatic and caustic masses, respectively. Similarly,
\(\delx\) and \(\delc\) represent the intrinsic scatter between the real
mass and the hydrostatic and caustic masses, respectively.

Weak priors were chosen for the model parameters. For each cluster,
the logarithmic masses \((\mu,\muX,\muC)\) were assigned a uniform
probability covering the range \(12:17\). The logarithmic bias terms
\((\kapx,\kapc)\) were assigned normal priors with mean 0 and standard
deviation 1 (roughly speaking, we believe the mass proxies to be
biased high or low by up to a factor of 10). The intrinsic scatter
terms were assigned normal priors (truncated at zero) with mean \(0.09\)
and standard deviation \(2.2\) (in natural log space this corresponds to
a mean of 0.2 and standard deviation of 5; a weak prior centred on a
scatter of \(20\%\)).

With this model, we can use our observations of
\((\muXh,\sigx,\muCh,\sigc)\) for each cluster to constrain
\((\kapx,\delx,\kapc,\delc)\) for the full sample. It is clear that the
pairs \((\kapx,\kapc)\) and \((\delx,\delc)\) will be highly degenerate,
but the mean bias between X-ray and caustic masses
\begin{align}
\kappa = \kapx - \kapc = \muX - \muC = \log_{10}\left(\frac{\MX}{\MC}\right)
\end{align}
and the intrinsic scatter between X-ray and caustic masses
\begin{align}
\delta = \sqrt{\delx^2+\delc^2}
\end{align}
will be constrained by the data.

The model was implemented in the probabilistic programming language
\emph{Stan} using the \emph{RStan}
interface\footnote{\url{http://mc-stan.org}},
and the parameters were sampled with 4 chains of \(5,000\) steps. This
procedure was repeated using the masses measured within different
radii to produce profiles of the mean bias between hydrostatic and
caustic masses.

It is useful to express the mean bias \(\kappa\) in terms of the mean
ratio \(\MX/\MC\). These are related by \(\kappa=\log_{10}(\MX/\MC)\). As
\(\kappa\) is normally distributed, the posterior distribution of
\(\MX/\MC\) is lognormal. We summarise this posterior of \(\MX/\MC\) by
quoting its median with errors given by the difference between the
median and 16th and 84th percentiles. Similarly, the posterior
distribution of \(\delta\) is found to be approximately lognormal, so we
also summarise this parameter by quoting its median with errors given
by the 16th and 84th percentiles.

\section{Results}
\label{sec:orgheadline7}
The caustic and hydrostatic cumulative mass profiles are shown for each cluster
in Figs. \ref{fig.xcprofs1} and \ref{fig.xcprofs2} in the appendix.
The hydrostatic mass profile of A1835 shows an un-physical declines at around $\rf$\footnote{The notation $\rf$ refers to the radius within which the mean density is 500 times the critical density at the cluster redshift. $M_{500}$ then refers to the mass enclosed by that radius.}. This was first reported in \citet{bon13}, and is interpreted as being due to the failure of the assumption of hydrostatic equilibrium at large radii.

\begin{figure}
\centering
\includegraphics[angle=0,width=240px]{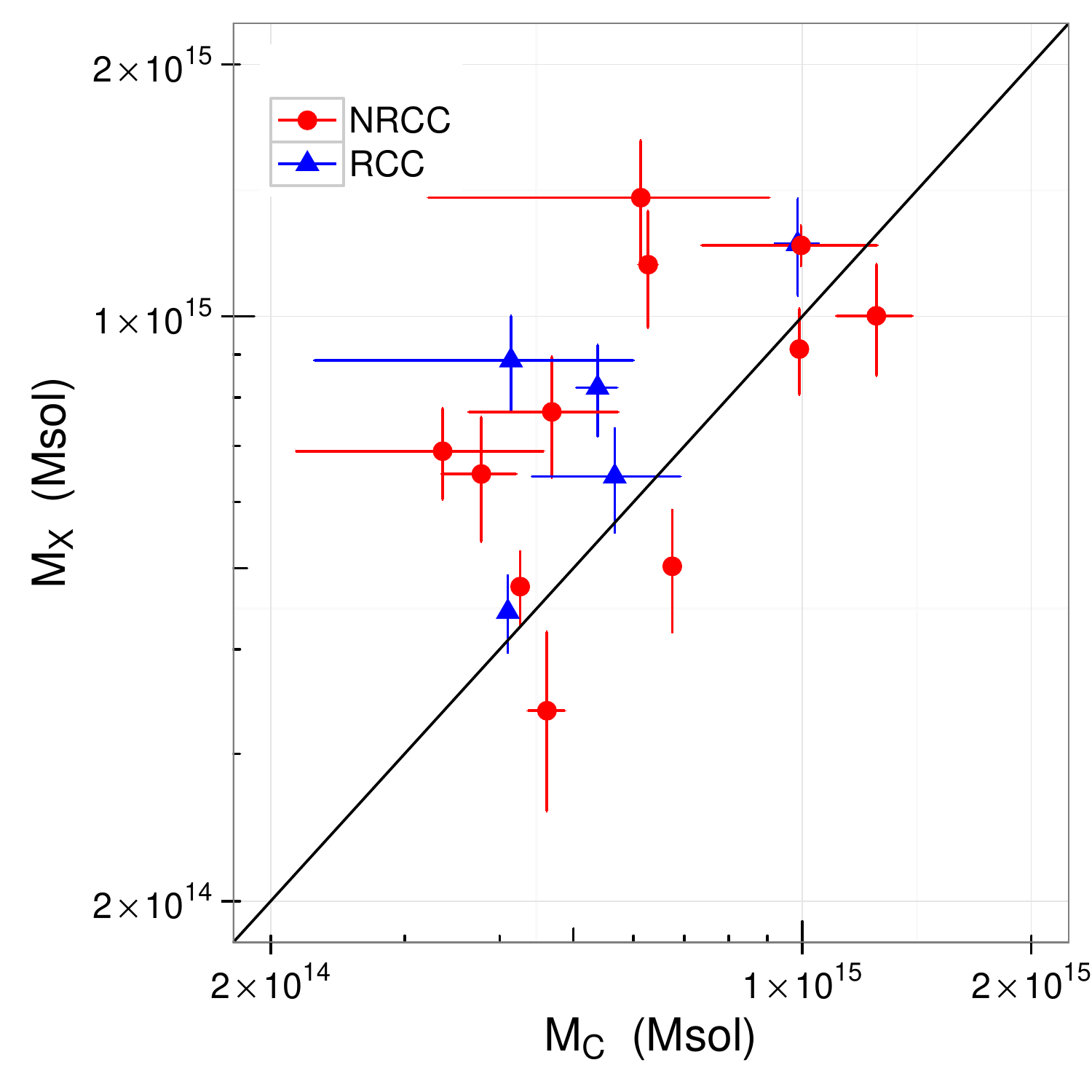}
\caption{\label{fig:orgparagraph1}
Comparison of hydrostatic (\(\MX\)) and caustic (\(\MC\)) masses, with both measured with the radius \(\rf\) defined from the hydrostatic mass profile. Points are coloured to indicate RCC (blue triangles) and NRCC (red circles) clusters. The solid line is the line of equality.}
\end{figure}

Using these profiles, the hydrostatic and caustic values of \(\mf\) were
then computed for each cluster within the radius \(\rf\) defined from
the hydrostatic mass profile. The resulting masses are compared in
Fig. \ref{fig:orgparagraph1} and summarised in Table \ref{tab.m500}. For our main
results we always compare quantities measured within the radius \(\rf\)
defined from the hydrostatic mass profiles. We note that this
introduces a covariance between the mass measurements, but we will see
below that fully consistent results are obtained when quantities are
measured in a fixed aperture of \(1\Mpc\).

% latex table generated in R 3.2.4 by xtable 1.8-2 package
% Tue May 17 15:15:58 2016
\begin{table}
  \centering
  \begin{tabular}{lclccc}
    \hline
    Cluster & z & Status & $\rf$ & $\MX$ & $\MC$ \\
            &   &        & Mpc & $10^{14}\Msol$ & $10^{14}\Msol$ \\
    \hline
    A0267   & 0.230 & NRCC & 0.99 & $3.4  \pm 0.8$ & $4.6  \pm 0.3$ \\
    A0697   & 0.282 & NRCC & 1.55 & $13.9 \pm 2.4$ & $6.1  \pm 2.9$ \\
    A0773   & 0.217 & NRCC & 1.38 & $9.1  \pm 1.1$ & $9.9  \pm 0.1$ \\
    A0963   & 0.206 & NRCC & 1.12 & $4.8  \pm 0.5$ & $4.26  \pm 0.04$ \\
    A1423   & 0.213 & RCC  & 1.09 & $4.4  \pm 0.5$ & $4.10  \pm 0.07$ \\
    A1682   & 0.234 & NRCC & 1.13 & $5.0  \pm 0.8$ & $6.74  \pm 0.04$ \\
    A1763   & 0.223 & NRCC & 1.42 & $10.0 \pm 1.5$ & $12.5 \pm 1.4$ \\
    A1835   & 0.253 & RCC  & 1.51 & $12.2 \pm 1.6$ & $9.9  \pm 0.7$ \\
    A1914   & 0.171 & NRCC & 1.52 & $11.5 \pm 1.8$ & $6.3  \pm 0.2$ \\
    A2111   & 0.229 & NRCC & 1.23 & $6.5  \pm 1.1$ & $3.8  \pm 0.4$ \\
    A2219   & 0.230 & NRCC & 1.52 & $12.2 \pm 0.7$ & $10.0 \pm 2.6$ \\
    A2261   & 0.224 & NRCC & 1.26 & $6.9  \pm 0.9$ & $3.4  \pm 1.2$ \\
    A2631   & 0.278 & NRCC & 1.28 & $7.7  \pm 1.3$ & $4.7  \pm 1.0$ \\
    RXJ1720 & 0.164 & RCC  & 1.36 & $8.2  \pm 1.0$ & $5.4  \pm 0.3$ \\
    RXJ2129 & 0.235 & RCC  & 1.22 & $6.4  \pm 0.9$ & $5.7  \pm 1.3$ \\
    Z3146   & 0.291 & RCC  & 1.34 & $8.9  \pm 1.2$ & $4.1  \pm 1.9$ \\
    \hline
  \end{tabular}
  \caption{Summary of the hydrostatic ($\MX$) and caustic ($\MC$) masses within the radius $\rf$ determined from the hydrostatic mass profile, given in column 4. The status column indicates the clusters' dynamical classification. \label{tab.m500}}
\end{table}

\begin{figure}
\centering
\includegraphics[angle=0,width=230px]{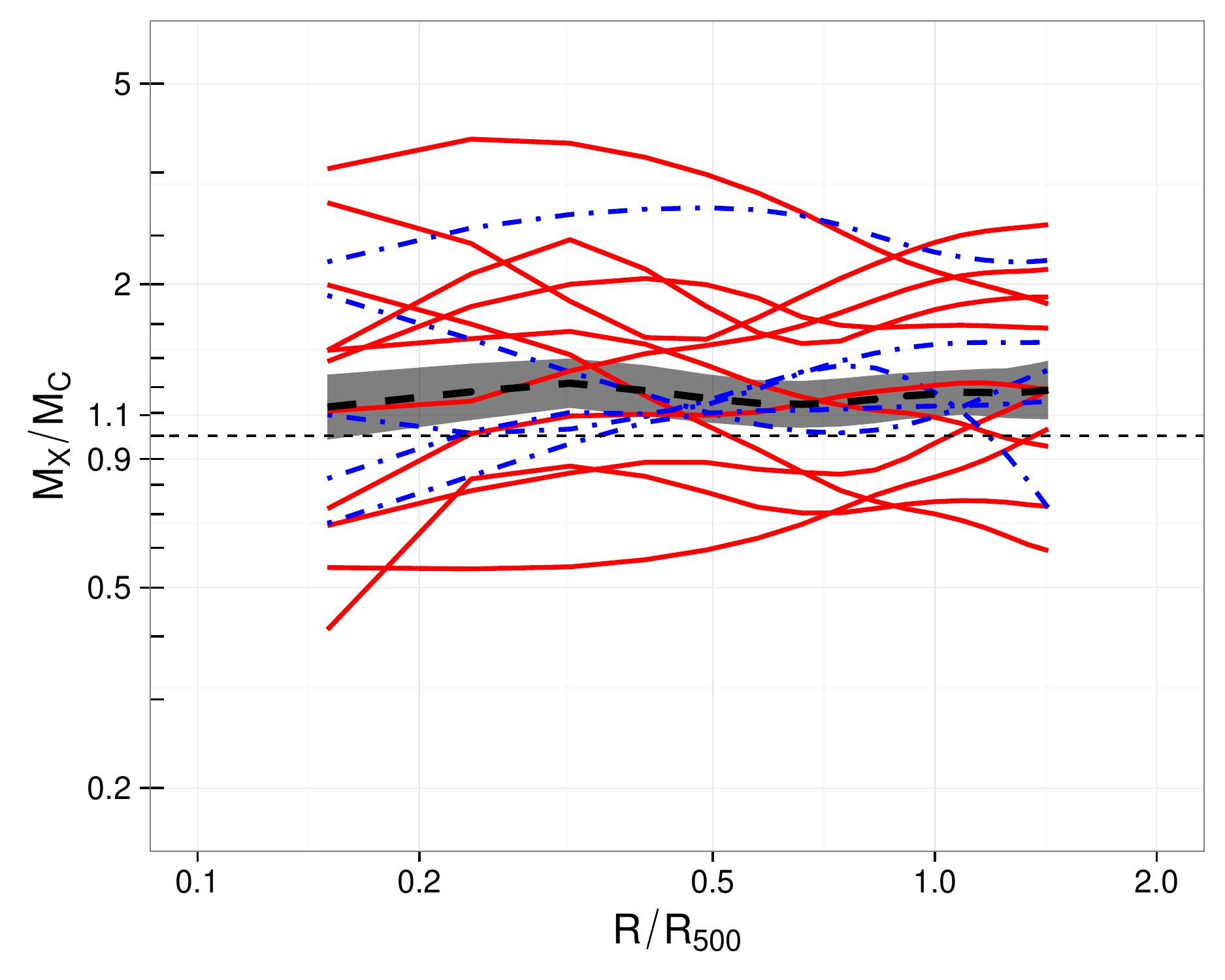}
\caption{\label{fig:orgparagraph2}
Profiles of the ratio of hydrostatic to caustic mass profiles for the sample. The ratios were computed in log space as described in the text. The mass profiles were scaled to the radius \(\rf\) determined from the hydrostatic mass profile for each cluster before fitting the bias at each radius. Lines are styled to indicate RCC (blue, dot-dashed) and NRCC (red, solid) clusters. The dashed black line shows the best fitting mean bias between hydrostatic and caustic mass, with the shaded region enclosing the \(1\sigma\) uncertainty.}
\end{figure}

\begin{figure}
\centering
\includegraphics[angle=0,width=230px]{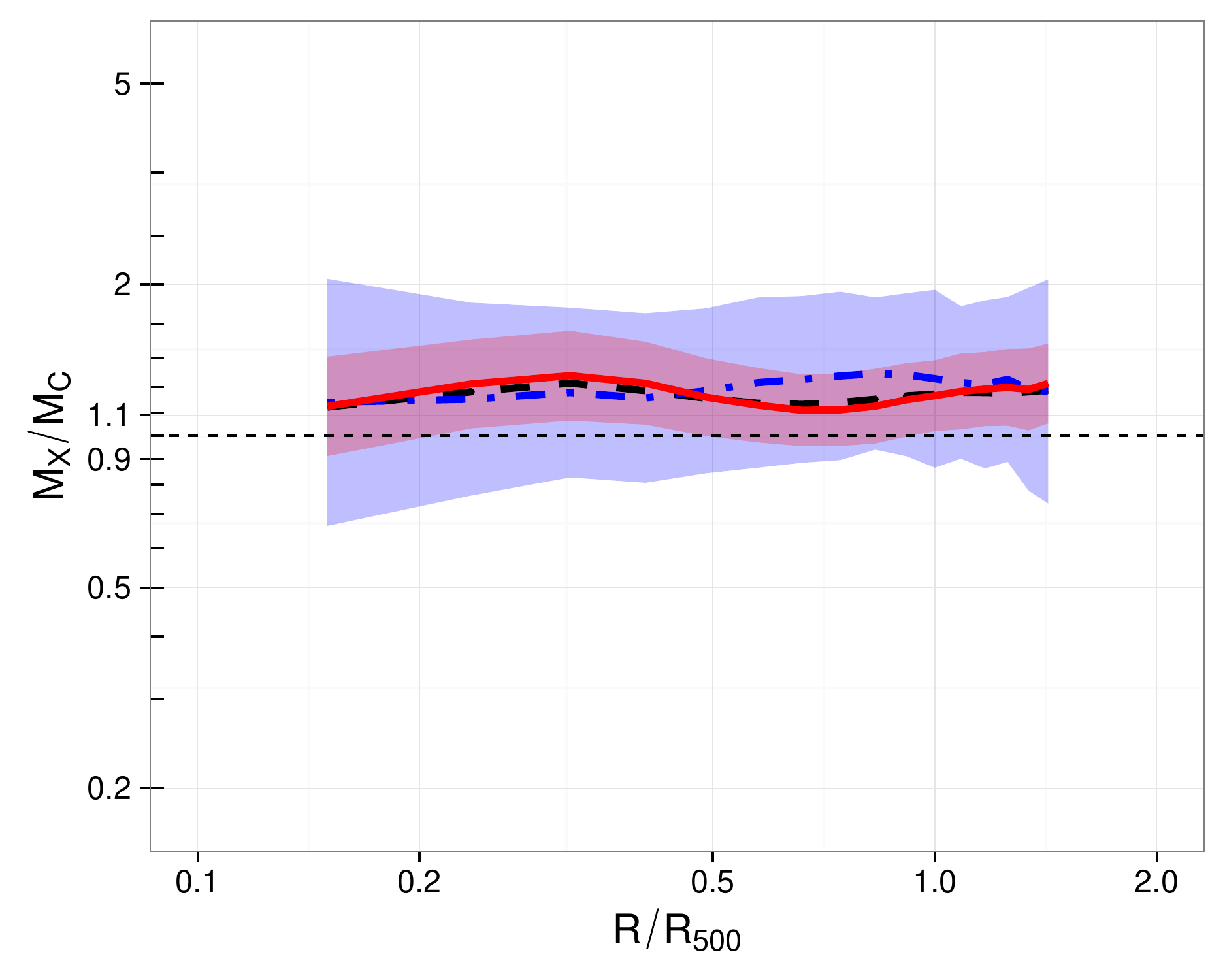}
\caption{\label{fig:orgparagraph3}
Profiles of the best-fitting mean bias between hydrostatic and caustic masses. The mass profiles were scaled to the radius \(\rf\) determined from the hydrostatic mass profile for each cluster before fitting the bias. The dashed black line shows the mean bias profile for the whole sample, while the blue (dot-dashed) and red (solid) lines with shaded error regions show the mean bias profiles for the RCC and NRCC subsets respectively.}
\end{figure}

Fig. \ref{fig:orgparagraph2} shows the observed \(\MX/\MC\) profile of each cluster
(computed as \(\muXh-\muCh\)), colour-coded to indicate if a cluster is
classified as RCC or NRCC. Also plotted is the profile of the mean
bias \(\kappa\) (expressed as \(\MX/\MC\) on this logarithmic plot). The
caustic and hydrostatic mass profiles agree to within \(\approx20\%\)
(\(\kappa\lta0.08\)) across the radial range. In Fig. \ref{fig:orgparagraph3},
the mean bias profiles of the RCC and NRCC clusters are shown
separately. These profiles demonstrate a similarly good agreement
between caustic and hydrostatic mass profiles for the two dynamical
subsets (albeit with larger uncertainties); in both cases the
agreement is better than \(\approx 30\%\) (\(\kappa\lta0.12\)).

As indicated in Figs. \ref{fig.xcprofs1} and \ref{fig.xcprofs2}, X-ray
temperature profiles were measured directly close to, or beyond,
\(\rf\) for all clusters. Hydrostatic mass profiles are extrapolated
based on the best-fitting temperature profile model beyond the extent
of the temperature profile. The median extent of the temperature
profiles is \(1.25~\rf\). Profiles of the mass ratios beyond that
point are less robust.

\begin{figure}
\centering
\includegraphics[angle=0,width=230px]{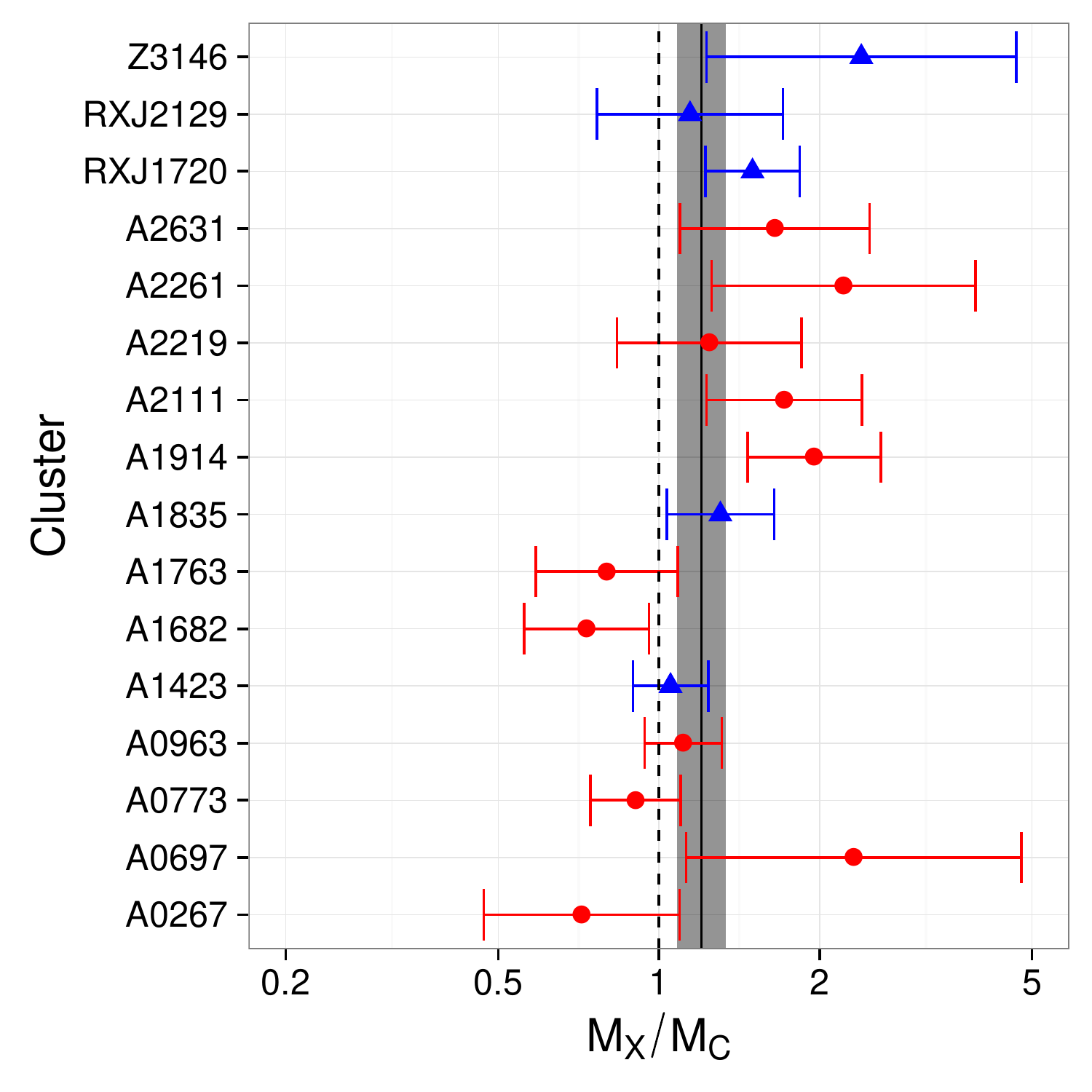}
\caption{\label{fig:orgparagraph4}
Ratio of hydrostatic to caustic masses at the hydrostatic \(\rf\) for each cluster (computed as \(\muXh-\muCh\)). Points are coloured to indicate RCC (blue triangles) and NRCC (red circles) clusters. The solid line and shaded band shows the mean bias between the hydrostatic and caustic masses and its uncertainty.}
\end{figure}

In Fig. \ref{fig:orgparagraph4}, the ratio of the hydrostatic to caustic masses at
the radius \(\rf\) determined from the hydrostatic mass profile is shown
for each cluster. Again, the ratios are computed as \((\muXh-\muCh)\).
At this radius, the two mass estimators agree well, with
\(\kappa=0.080\pm0.046\), corresponding to
\(\MX/\MC=1.20^{+0.13}_{-0.11}\). The intrinsic scatter between the
hydrostatic and caustic mass estimators at this radius is
\(\delta=0.11\pm0.05\dex\), corresponding to an intrinsic scatter of
\(23^{+13}_{-10}\%\). The RCC and NRCC subsamples show consistent
results, albiet with weaker constraints.

These summary statistics are captured in Table \ref{tab:orgtable2}, along with
the same quantities derived in a fixed aperture of \(1\Mpc\) for each
cluster. The parameter values are insensitive to the choice of
aperture, demonstrating that our results are not significantly
influenced by scaling the caustic profiles to the hydrostatic estimate
of \rf.

\renewcommand{\arraystretch}{1.4}
\begin{table}
\caption{\label{tab:orgtable2}
Summary statistics of the ratio of the mean bias between hydrostatic and caustic masses, measured in the hydrostatic estimate of \(\rf\) and in a fixed aperture of \(1\Mpc\). The second column gives the subset being considered and the third column gives the mean bias. The mean bias is expressed as the mean ratio of masses in column four, and the intrinsic scatter between the mass estimators (converted to a percentage) is given in column five. Note that the quantities in columns four and five are summaries of a lognormal distribution using a median and percentiles as described in \textsection \ref{sec:orgheadline5}.}
\centering
\begin{tabular}{llccc}
\hline
Aperture & Subset & \(\kappa\) & \(\MX/\MC\) & \(\delta\) (\%)\\
\hline
\rf & All & \(0.080\pm0.046\) & \(1.20^{+0.13}_{-0.11}\) & \(23^{+13}_{-10}\)\\
\rf & RCC & \(0.12\pm0.16\) & \(1.31^{+0.38}_{-0.28}\) & \(41^{+51}_{-22}\)\\
\rf & NRCC & \(0.071\pm0.073\) & \(1.17^{+0.20}_{-0.16}\) & \(36^{+20}_{-17}\)\\
\hline
\(1\Mpc\) & All & \(0.062\pm0.046\) & \(1.15^{+0.13}_{-0.10}\) & \(27^{+12}_{-09}\)\\
\hline
\end{tabular}
\end{table}

\section{Discussion}
\label{sec:orgheadline11}
\subsection{Biases in hydrostatic and caustic masses}
\label{sec:orgheadline8}
We have performed one of the first comparisons between mass profiles
of galaxy clusters determined from their hot gas via hydrostatic
assumptions, and the dynamics of their galaxies via the caustic
method. These two methods are completely independent, and are subject
to different assumptions and systematics. We found that, while
significant scatter is present between the two estimators, the average
agreement is good. As demonstrated in Fig. \ref{fig:orgparagraph2}, the masses
agree to better than \(\approx20\%\) on average over the full radial
range sampled by both techniques. Importantly, neither of the mass
measurement techniques assumed a functional form for the mass profile
(although the hydrostatic analysis did use parametric temperature and
density profiles). The agreement we find is not a consequence of a
common parametrisation of the mass profile.

This good average agreement is somewhat surprising given that various
observational and theoretical studies have suggested that hydrostatic
and caustic masses are biased in opposite directions at around \(\rf\).
The caustic mass estimates make the assumption that the filling factor
\citep[a quantity related to the ratio of the mass gradient to the gravitational potential; see][for details]{ser11}
is constant with radius. N-body simulations have shown this
approximation breaks down in the inner parts of clusters, and that the
caustic technique will tend to overestimate the true mass by
\(\sim10-20\%\) at \(\rf\), increasing to smaller radii \citep{ser11}.
Meanwhile, hydrodynamical simulations and observational comparisons
with weak lensing masses indicate that hydrostatic masses could be
biased low by up to \(\sim30\%\) at \(\rf\) due to the presence of
non-thermal pressure in the gas
\citep[e.g.][]{lau09,ras12,mah13,von14,nel14a}.

For both caustic and hydrostatic masses, the effect of these expected
biases should lead to \(\MX/\MC<1\). The observed mean ratio of
\(1.20^{+0.13}_{-0.11}\) at \(\rf\) places a \(\sim10\%\) upper limit (at
\(3\sigma\)) on the combination of these two systematics (i.e.
\(\MX/\MC\gta0.9\) at \(\approx3\sigma\)).

Our work provides a valuable comparison with recent studies that
have attempted to constrain the level of hydrostatic bias by comparing
hydrostatic and weak lensing masses. At face value there appear to be
some large discrepancies, with e.g. Weighing the Giants
\citep[WtG;][]{von14} and the Canadian Cluster Cosmology Project
\citep[CCCP;][]{hoe15} finding hydrostatic masses to be biased low by
\(\approx25-30\%\) at \(\rf\), while \citet{gru14}, \citet{isr14}, \citet{app16} and
\citet{smi16} found ratios of hydrostatic to lensing mass (\(\MX/\MWL\))
in the range \(0.92-1.06\) at \rf, which were all consistent with zero
hydrostatic bias. \citet{smi16} limited the WtG and CCCP samples to
clusters with \(z<0.3\) and recomputed biases using methods consistent
with their own. In this analysis the WtG and CCCP measurements became
consistent with a \(\lta 10\%\) bias in the hydrostatic masses towards
lower values (and consistent with zero bias). While it is not yet
clear which of the different analysis methods used in these studies
was optimal, the lensing-based studies appear overall to be
consistent with a low or zero value of hydrostatic bias at \rf, at
least for clusters at \(z<0.3\).

Our measurement of \(\MX/\MC=1.20^{+0.13}_{-0.11}\) provides significant
support for low or zero hydrostatic bias, in a way that is independent
of any systematics affecting lensing-derived masses. Further support
comes from \citet{rin16} who compared velocity dispersions and masses
derived from the Sunyaev-Zel'dovich effect (the latter calibrated
from hydrostatic masses), again inferring no significant hydrostatic
bias.

It is interesting to note that the agreement we found between the
hydrostatic and caustic mass profiles does not appear dependent on the
X-ray morphology, with the mean mass ratio profiles of the RCC and
NRCC clusters in good agreement across the radial range probed (Fig.
\ref{fig:orgparagraph3}). Any discussion of this agreement is necessarily
limited by the large uncertainties on the profiles for these subsets,
but the results are suggestive that any non-thermal pressure effects
are present at similar levels in the most relaxed clusters and the
rest of the sample. The precision of this result is primarily
limited by the small number of RCC clusters in the present sample, and
will be investigated in more detail when the analysis is extended to
the full flux-limited HeCS sample of 50 clusters. The inferred
similarity of the hydrostatic bias for relaxed and unrelaxed clusters
agrees qualitatively with the results from hydrodynamical simulations
which show a fairly modest difference in the level of hydrostatic bias
between relaxed and unrelaxed clusters at \(\rf\)
\citep[e.g.][]{nag07,lau09}.

The intrinsic scatter between the hydrostatic and caustic masses is
\(23^{+13}_{-10}\%\) at \(\rf\). This is similar to the expected
\(\sim30\%\) intrinsic scatter in caustic mass at a fixed true mass,
caused by projection effects for non-spherical clusters
\citep{ser11,gif13}. These projection can thus account for all of the
scatter between the mass estimators. Another possible contribution to
the scatter comes from the centering of the mass profiles. Due to the
hydrostatic and caustic analyses being performed independently, their
profiles were not centred on the same coordinates. For each cluster,
the X-ray profiles were centred on the centroid of the X-ray emission,
while the caustic profiles were centred on the hierarchical centre of
the galaxy distribution. This difference in central position should
not affect the average agreement between the mass profiles, but will
contribute to the intrinsic scatter between the masses.

\subsection{Possible systematic effects}
\label{sec:orgheadline9}
For our main results, we scaled the caustic mass profiles to the
hydrostatic estimate of \rf. Such scaling is useful, since \(\rf\) is a
commonly-used reference radius for mass comparisons, but it introduces
covariance between the masses. This could suppress scatter between the
two estimators. We verified that the use of the hydrostatic \(\rf\) did
not significantly influence our results by repeating the analysis
using unscaled profiles and comparing the masses at \(1\Mpc\). As shown
in Table \ref{tab:orgtable2} the results were insensitive to the choice of
radius, and the average agreement between the two mass estimators
remained good.

An additional systematic that can effect the hydrostatic masses is the
calibration of the X-ray observatories. It is well known that
\(\Chandra\) and \(\XMM\) show systematic differences in temperatures
measured for hot clusters. Recently, \citet{sch15} showed that
hydrostatic masses are on average \(14\%\) lower when inferred from
\(\XMM\) observations than from \(\Chandra\) \citep[but see][]{mar14}.
Thus, if we scaled our \(\Chandra\) hydrostatic masses to the \(\XMM\)
calibration, our inferred \(\MX/\MC\) would reduce to \(1.05\), more
easily accommodating a larger hydrostatic bias as inferred from some
weak lensing comparisons and/or the expected systematic overestimate
of the caustic masses at \(\rf\). However, it is by no means clear that
this is the correct approach. Firstly, the three imaging detectors on
\(\XMM\) do not measure consistent temperatures with each-other
\citep{sch15}. Secondly, calibrating the \(\XMM\) derived hydrostatic
mass scale used by \citet{pla14d} to the higher \(\Chandra\) masses
helps reduce the tension between the cosmological parameters inferred
from the \emph{Planck} cluster counts and the cosmic microwave background
\citep{sch15}. It is clear that the X-ray calibration is a significant
systematic uncertainty affecting the interpretation of our results.

\subsection{Direct comparisons of our masses with other work}
\label{sec:orgheadline10}
In this section we directly compare the masses measured for the
clusters in our sample with those from other work.

We compared our hydrostatic masses with those measured by other
authors using \emph{Chandra} observations of the same clusters. All of the
clusters in our sample were analysed by \citet{mar14}, and nine were
in the sample of \citet{mah13}. In both cases we remeasured our
hydrostatic masses within the same \(\rf\) radii used in the comparison
study, to ensure consistency. The weighted mean ratio of our
hydrostatic masses to those of \citet{mar14} was \(1.05\pm0.07\), and to
those of \citet{mah13} was \(1.04\pm0.09\); a very good agreement in
both cases. We can thus conclude that the measured \(\MX/\MC\) is
unlikely to be overestimated due to systematics in the X-ray analysis.

Many of the clusters in our sample have been studied by one or more
weak-lensing project \citep[e.g.][]{oka16,von14a,hoe15,mer15}. In
\citet{gel13}, mass profiles from caustics and weak lensing were
compared for 19 clusters (17 from HeCS, with lensing masses from
various sources). Caustic masses were found to be larger than lensing
masses at radii smaller than \(\rt\), and in good agreement around
\(\rt\). Since that comparison was made, however, many of the lensing
masses that were used have been revised upwards following updated
analyses \citep{oka16,hoe15}. \citet{hoe15} compared caustic and
lensing masses within \(\rt\) for 14 clusters in common between their
lensing sample and the HeCS. They found a mean ratio of lensing to
caustic masses of \(\MWL/\MC=1.22\pm0.07\). The difference from the good
agreement found at \(\rt\) by \citet{gel13} is at least partly due to
the revision upwards of the lensing masses in \citet{hoe15} compared
to those used in \citet{gel13}. Also, two of the clusters in the
comparison sample contain multiple clusters along the line of sight.
Because weak lensing measures the total mass of all systems while the
caustic technique measures the mass of the largest cluster, these mass
estimates are significant outliers and may bias the mean ratio
\citep{gel13,hoe15}.

A full comparison of our caustic and hydrostatic masses with the range
of new and updated lensing masses requires a careful comparison of
mass profiles on a cluster-by-cluster basis taking into account
contamination by foreground structures \citep[e.g.][]{hwa14}. This is
beyond the scope of the current paper. For the present, we performed a
simple comparison of our caustic and hydrostatic masses with the
lensing masses of \citet{hoe15} for nine clusters in common between
the samples
\citep[the][dataset was chosen for this simple comparison as it is recent and has the largest overlap with our current sample from a single lensing study]{hoe15}.
For this comparison, we used the NFW \(\mf\) masses from \citet{hoe15},
and recomputed the hydrostatic and caustic masses within the radius
\(\rf\) measured from the weak lensing data. We found that both the
X-ray and caustic masses were consistent, on average, with the lensing
masses, though the small number of clusters available limited the
precision of the comparison.

\section{Summary}
\label{sec:orgheadline12}
For 16 massive clusters, we compared the hydrostatic and caustic masses
based, respectively, on X-ray and optical data. We conclude:

\begin{itemize}
\item The hydrostatic and caustic masses agree to better than
\(\approx20\%\) on average across the radial range covered by both
techniques. The mass profiles were measured independently and the
agreement in masses not due to a shared parametrisation of the mass
profiles.

\item The ratio \(\MX/\MC\gta0.9\) at \(\rf\) (at \(3\sigma\)), placing a limit
on the amount by which hydrostatic masses are underestimated or
caustic masses are overestimated. Our results favour a low (or
zero) value of hydrostatic bias, consistent with some of the recent
lensing-based estimates.

\item There is no indication of any dependence of \(\MX/\MC\) on the X-ray
morphology of the clusters although the comparison is currently
limited by the small sample size, indicating that the hydrostatic
masses are not strongly systematically affected by the dynamical
state of the clusters.

\item The scatter between \(\MX\) and \(\MC\) is \(23^{+13}_{-10}\%\) at \(\rf\),
and is consistent with being due to the expected scatter in caustic
mass from projection effects.
\end{itemize}

We plan to use new \emph{Chandra} observations to extend this analysis to
the complete flux-limited sample of 50 HeCS clusters.

\section*{Acknowledgements}

BJM and PAG acknowledge support from STFC grants ST/J001414/1 and
ST/M000907/1. AD acknowledges support from the grant Progetti di
Ateneo/CSP$\_$TO$\_$Call2$\_$2012$\_$0011
``Marco Polo'' of the University of Torino, the INFN grant InDark, and
the grant PRIN 2012 ``Fisica Astroparticellare Teorica'' of the
Italian Ministry of University and Research.

\bibliographystyle{mnras}
\bibliography{clusters}

\appendix

\section{Mass profile plots}
\label{sec:orgheadline13}
\begin{figure*}
\centering
\includegraphics[angle=0,width=210px]{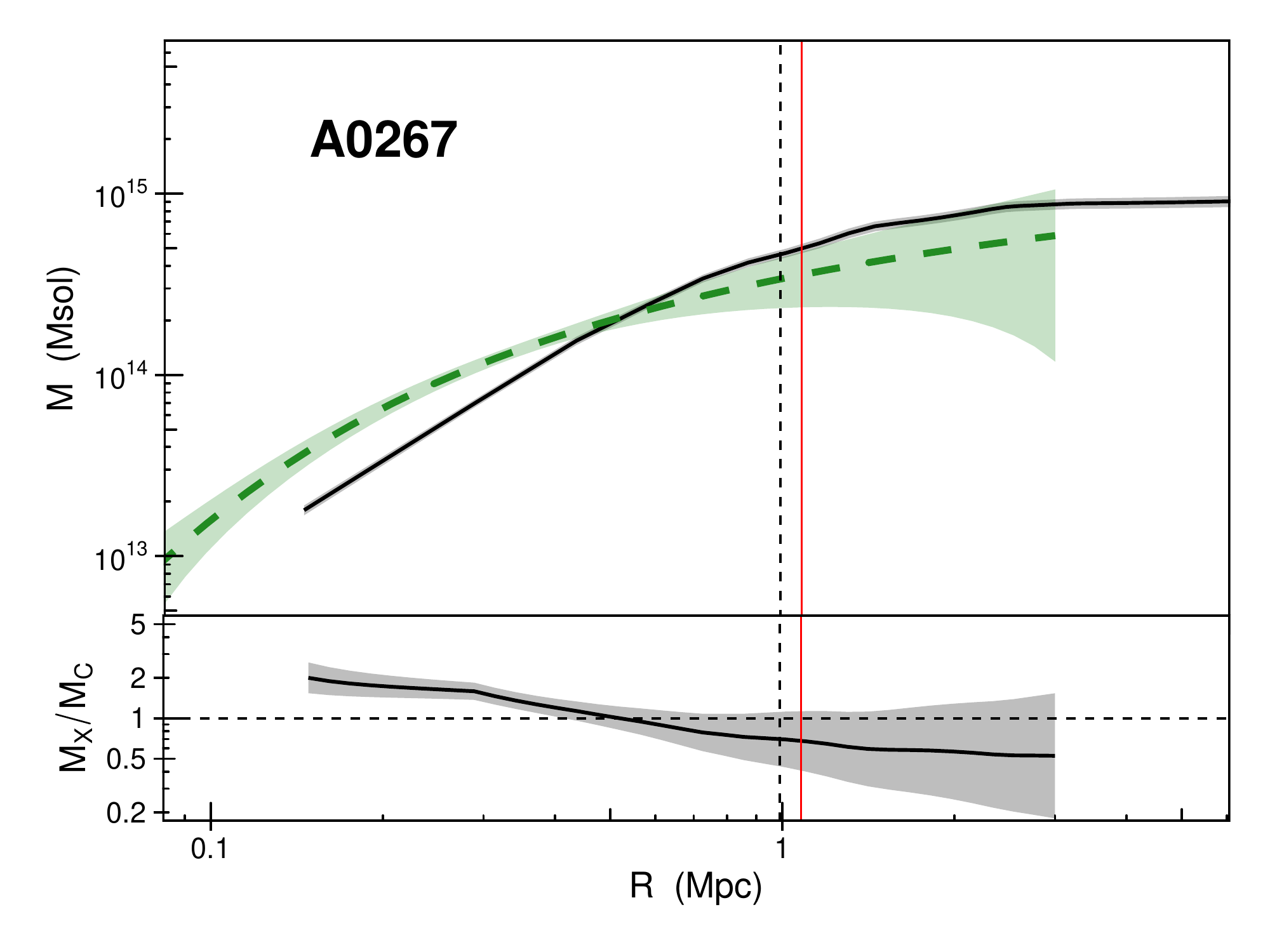} \qquad
\includegraphics[angle=0,width=210px]{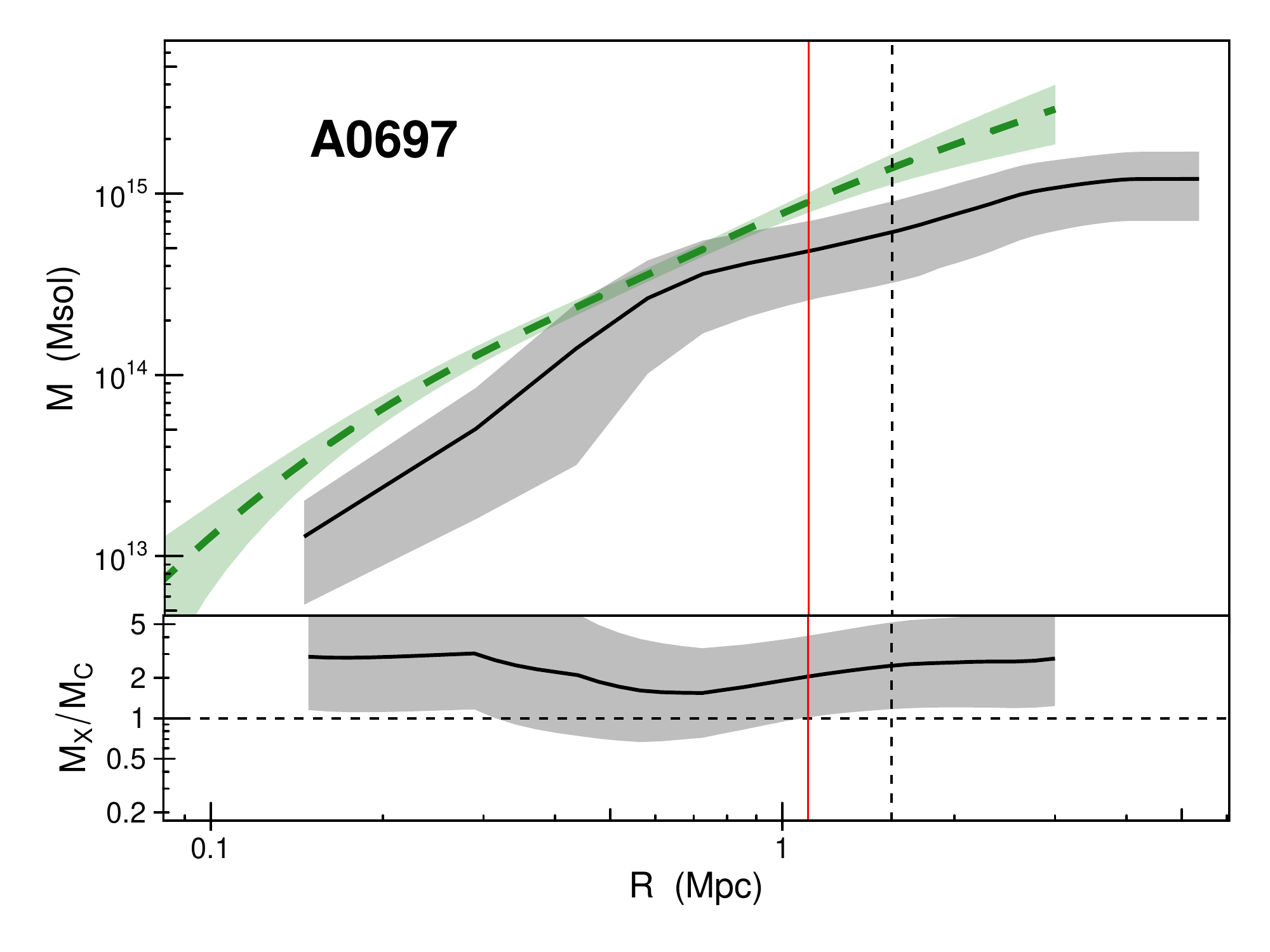} \\
\includegraphics[angle=0,width=210px]{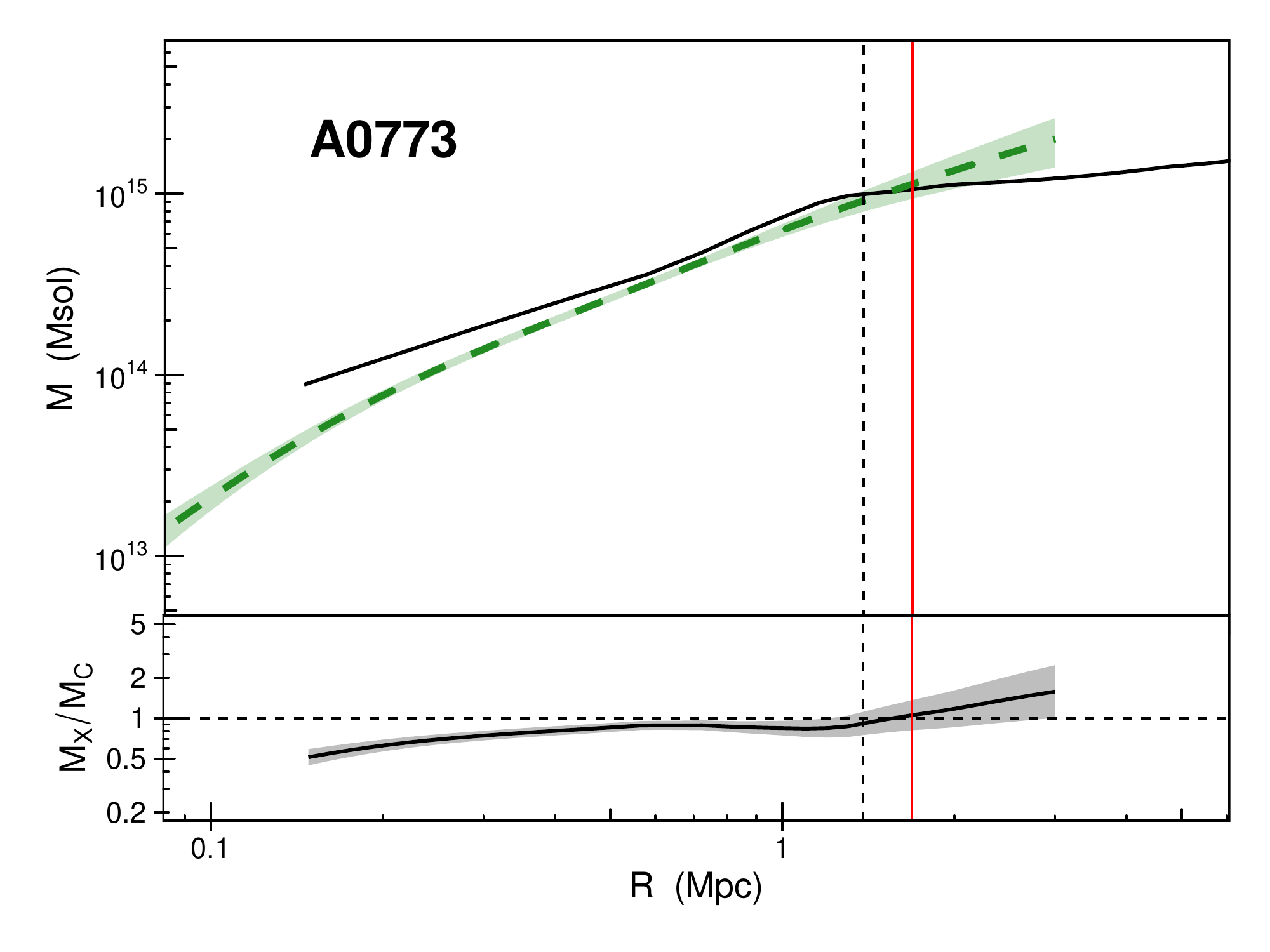} \qquad
\includegraphics[angle=0,width=210px]{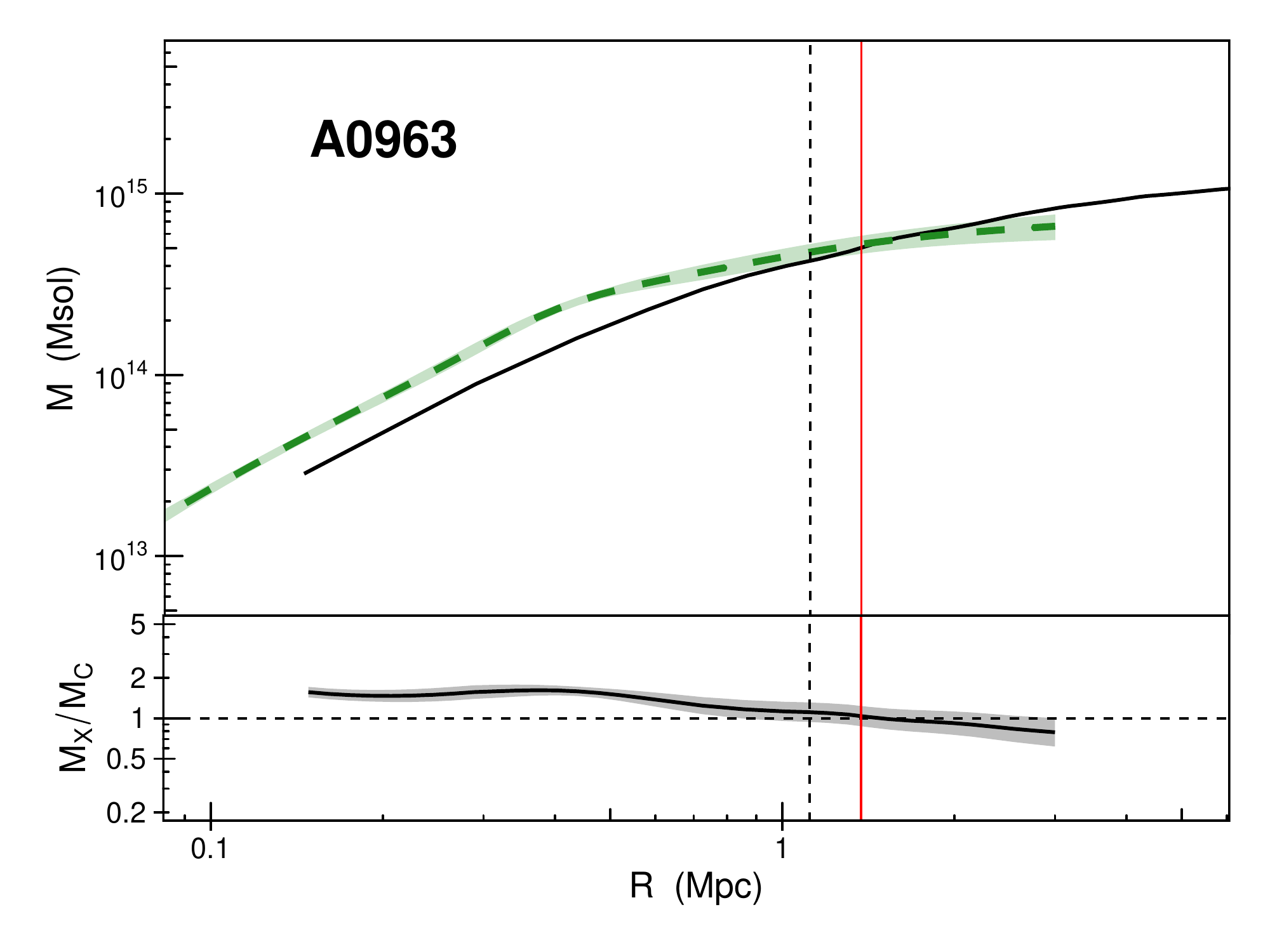} \\
\includegraphics[angle=0,width=210px]{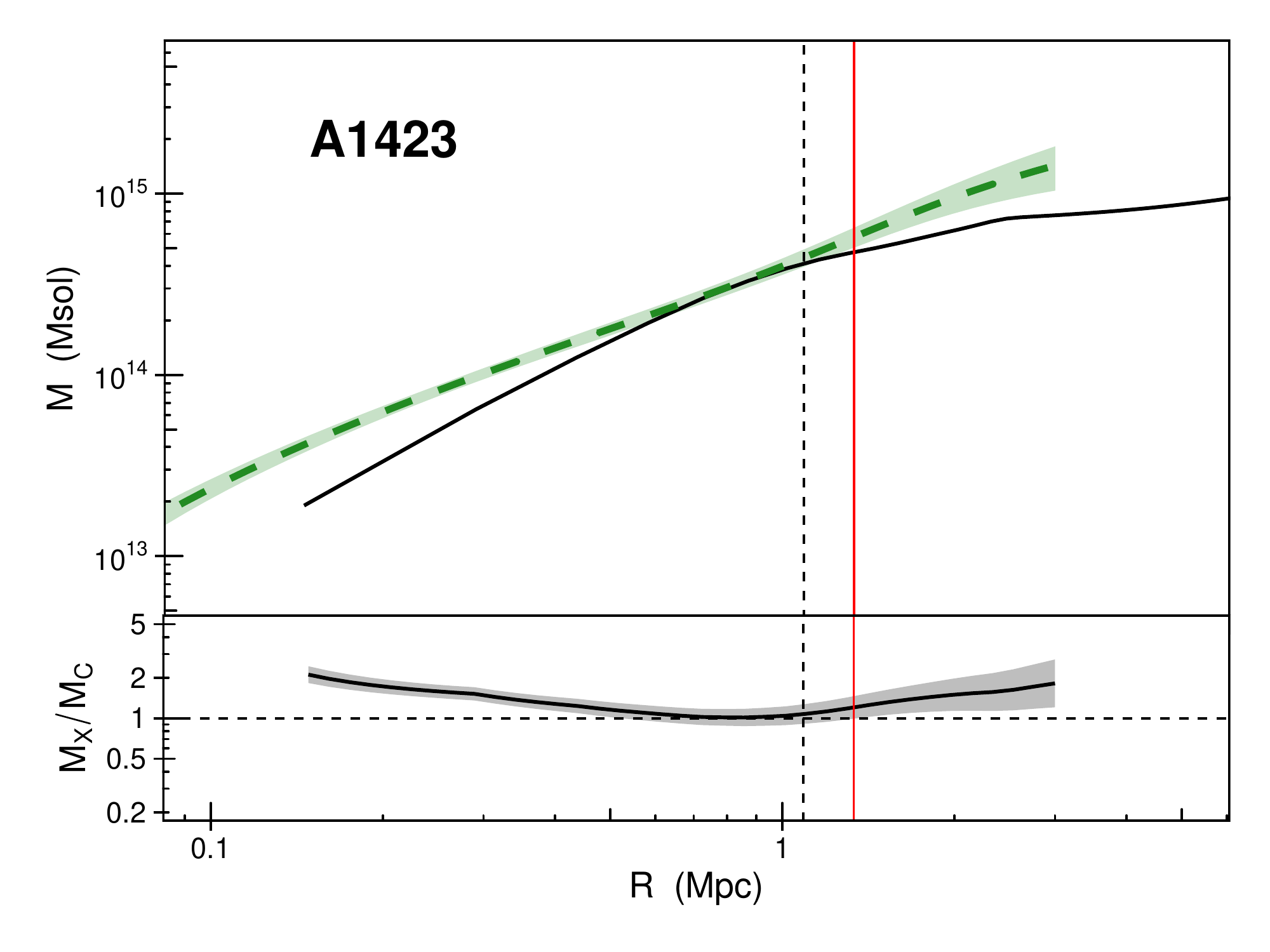} \qquad
\includegraphics[angle=0,width=210px]{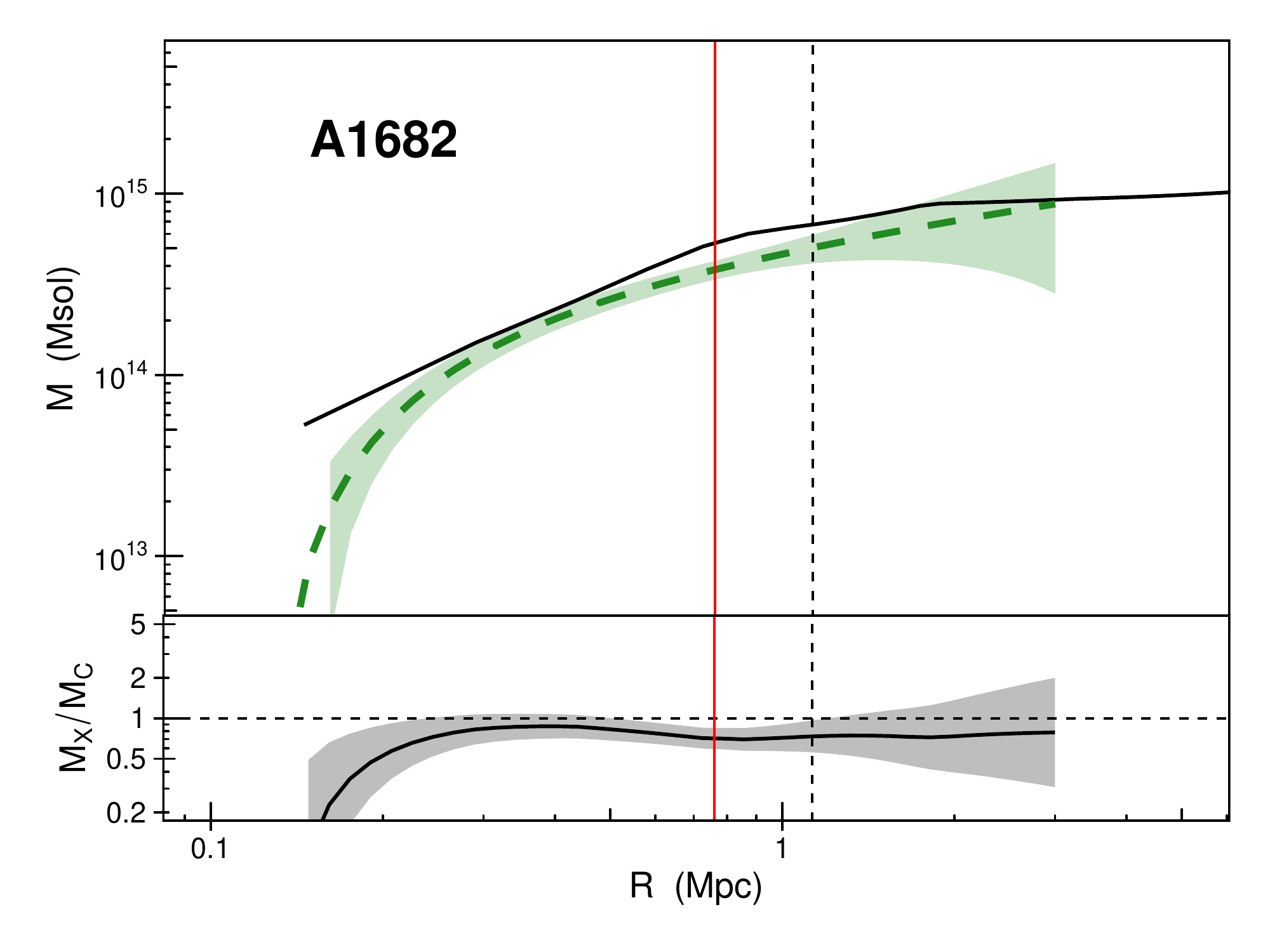} \\
\includegraphics[angle=0,width=210px]{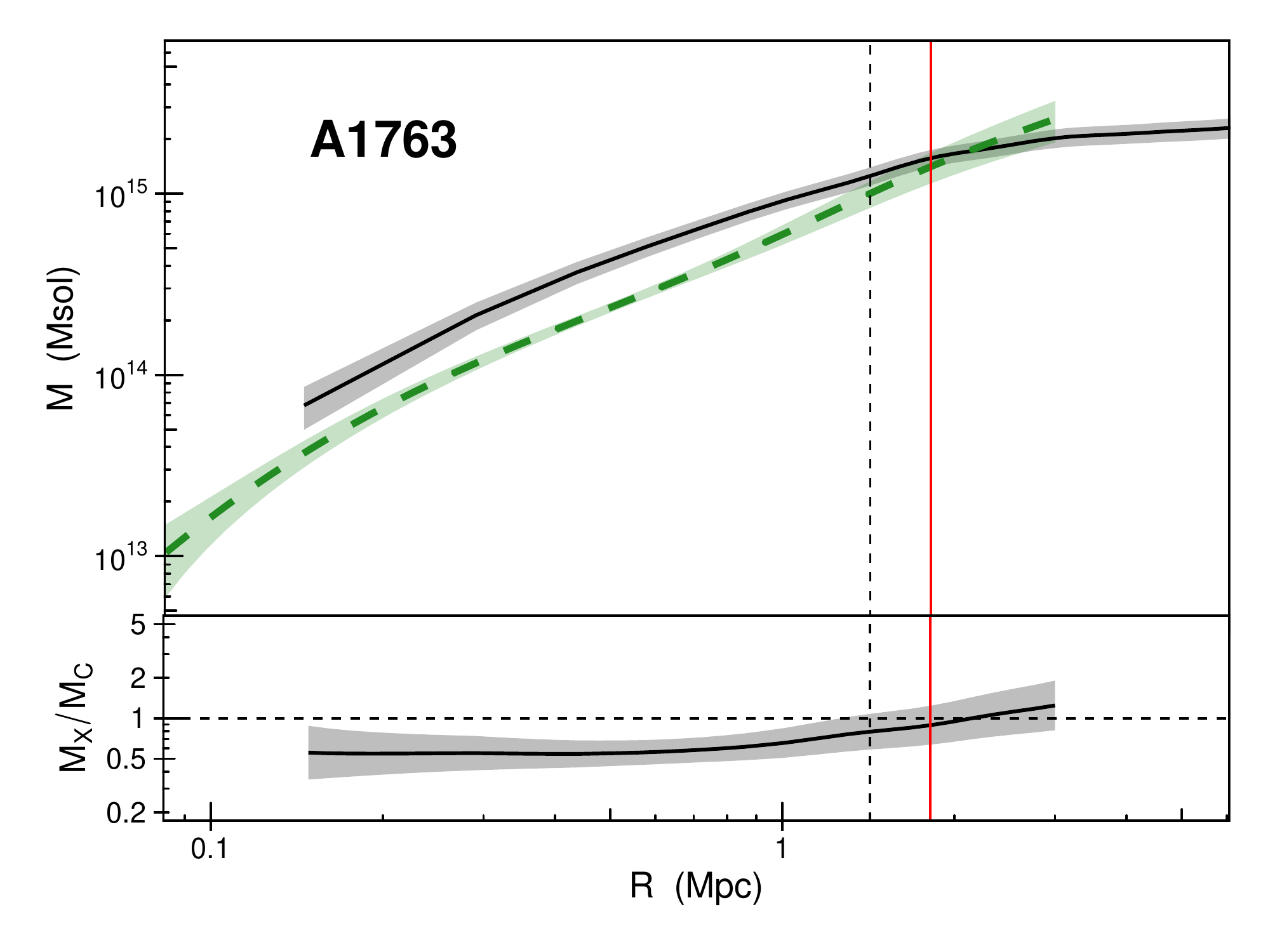} \qquad
\includegraphics[angle=0,width=210px]{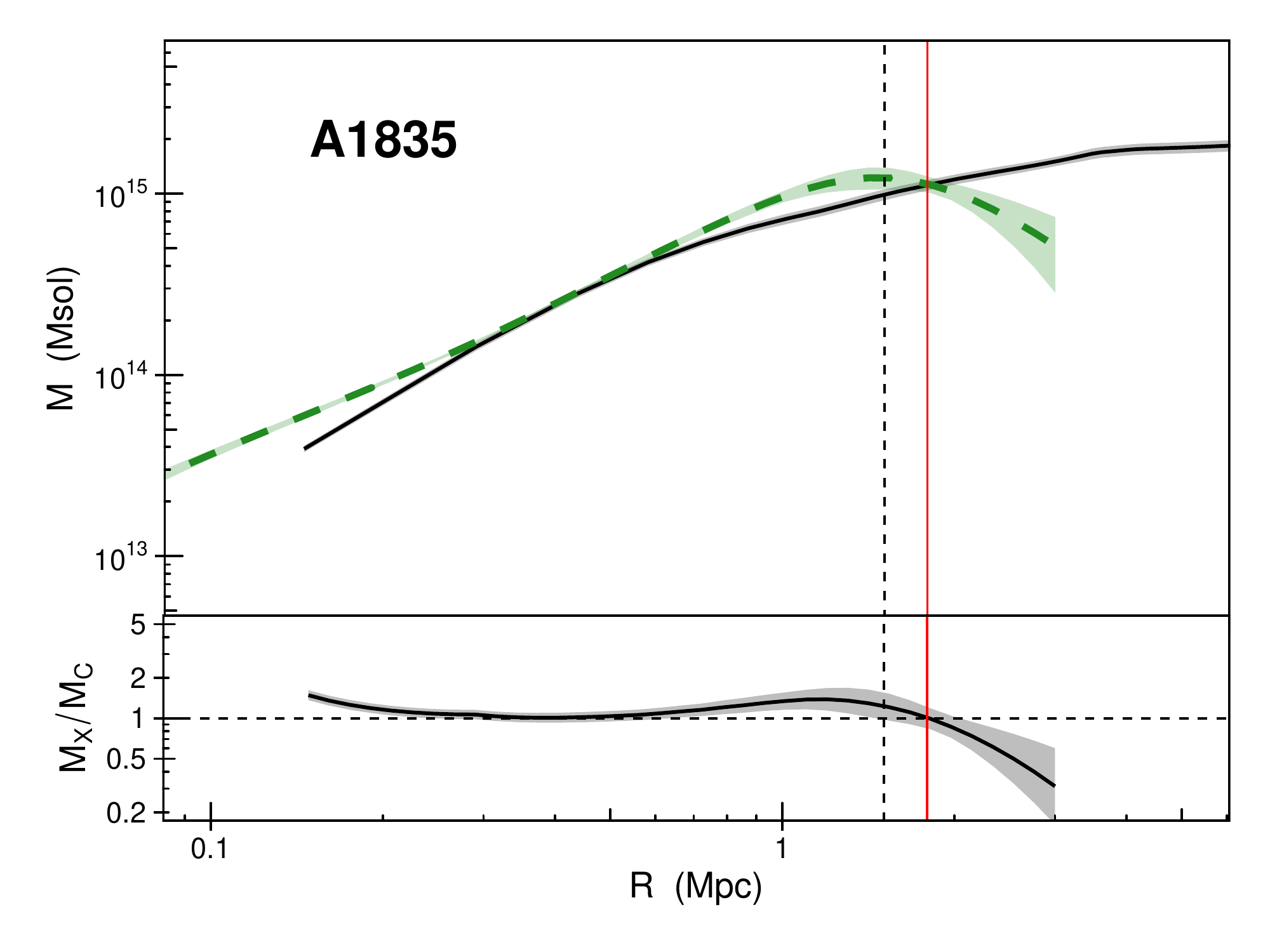}
\caption{\label{fig.xcprofs1} The upper panels in each plot show the caustic (black solid) and hydrostatic (green dashed) mass profiles for each cluster, while the lower panels show the ratio of the hydrostatic to caustic masses. The shaded regions indicate $1\sigma$ uncertainties. The vertical black dashed line indicates the value of $\rf$ estimated from the hydrostatic profile, and the vertical red solid line indicates the extent of the measured temperature profile - hydrostatic masses beyond this radius are based on extrapolation.}
\end{figure*}

\begin{figure*}
\centering
\includegraphics[angle=0,width=210px]{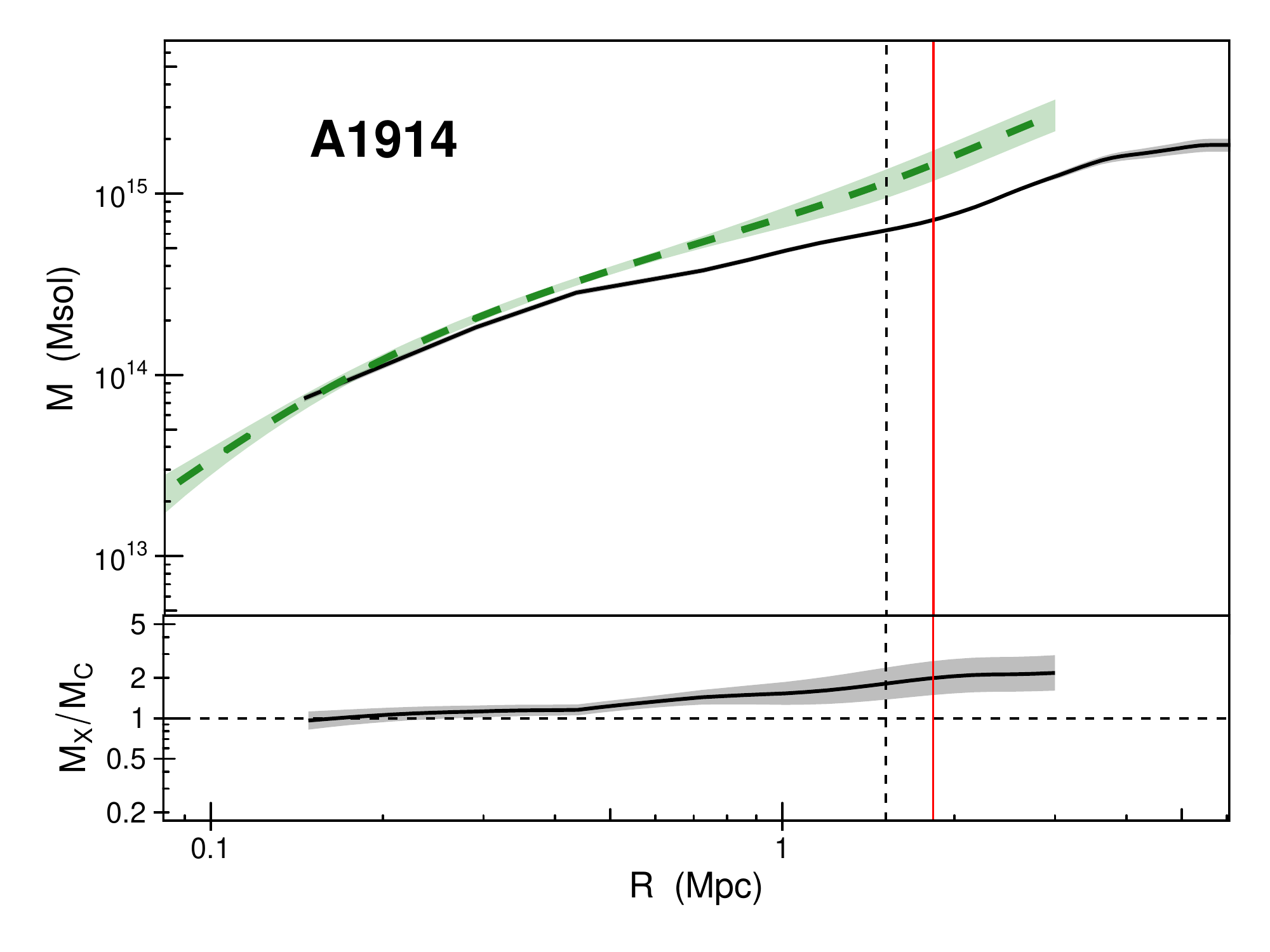} \qquad
\includegraphics[angle=0,width=210px]{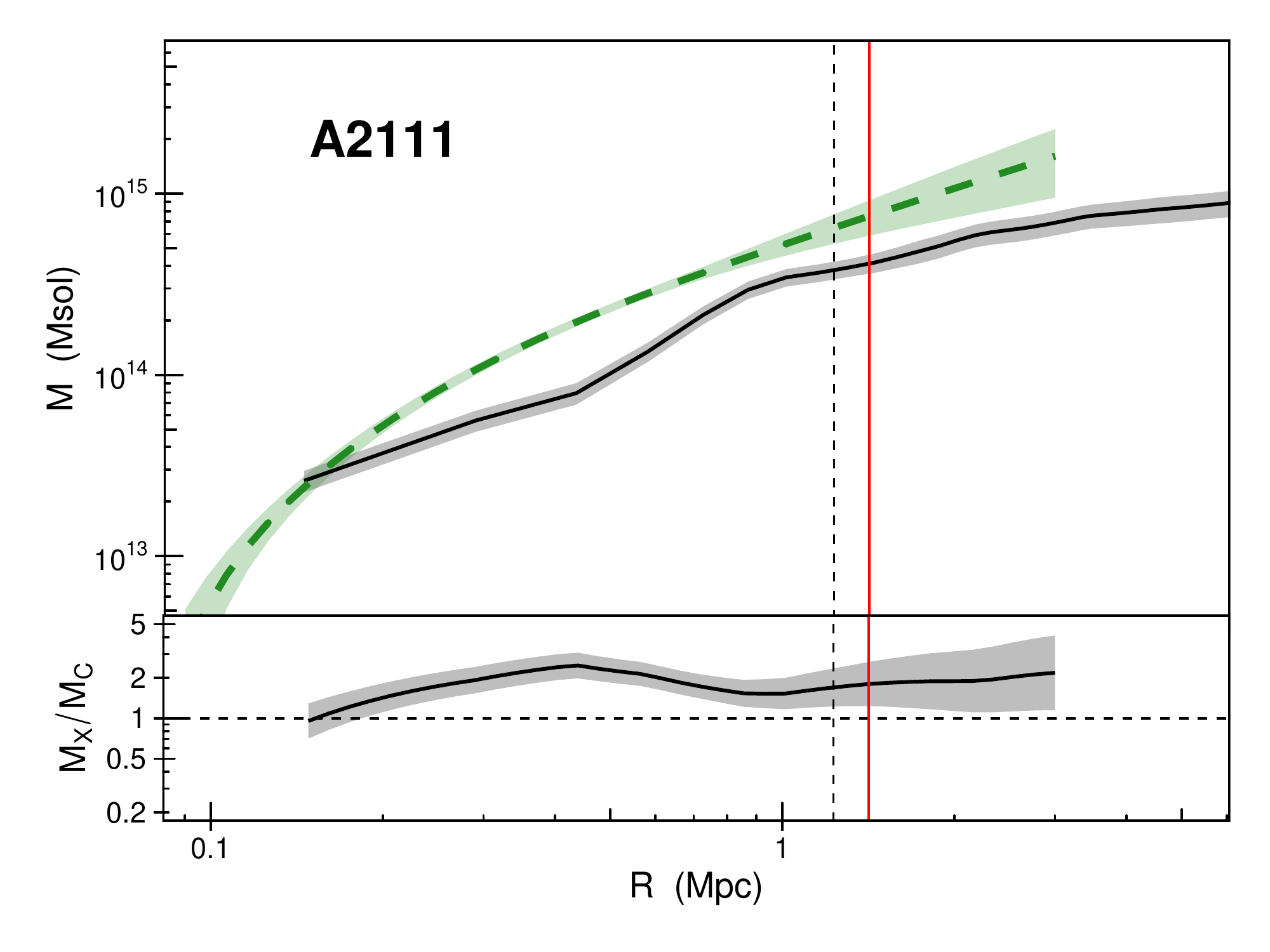} \\
\includegraphics[angle=0,width=210px]{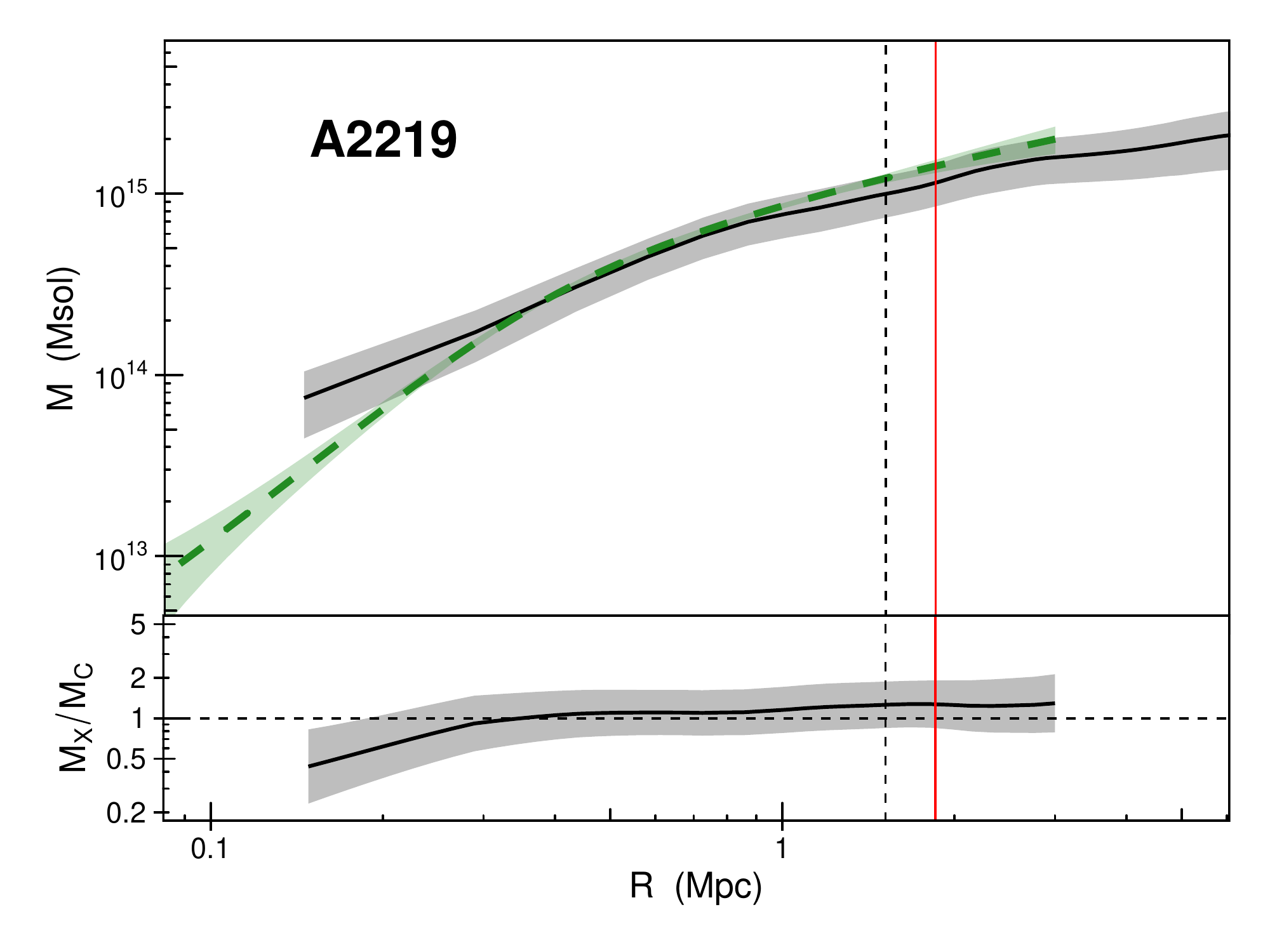} \qquad
\includegraphics[angle=0,width=210px]{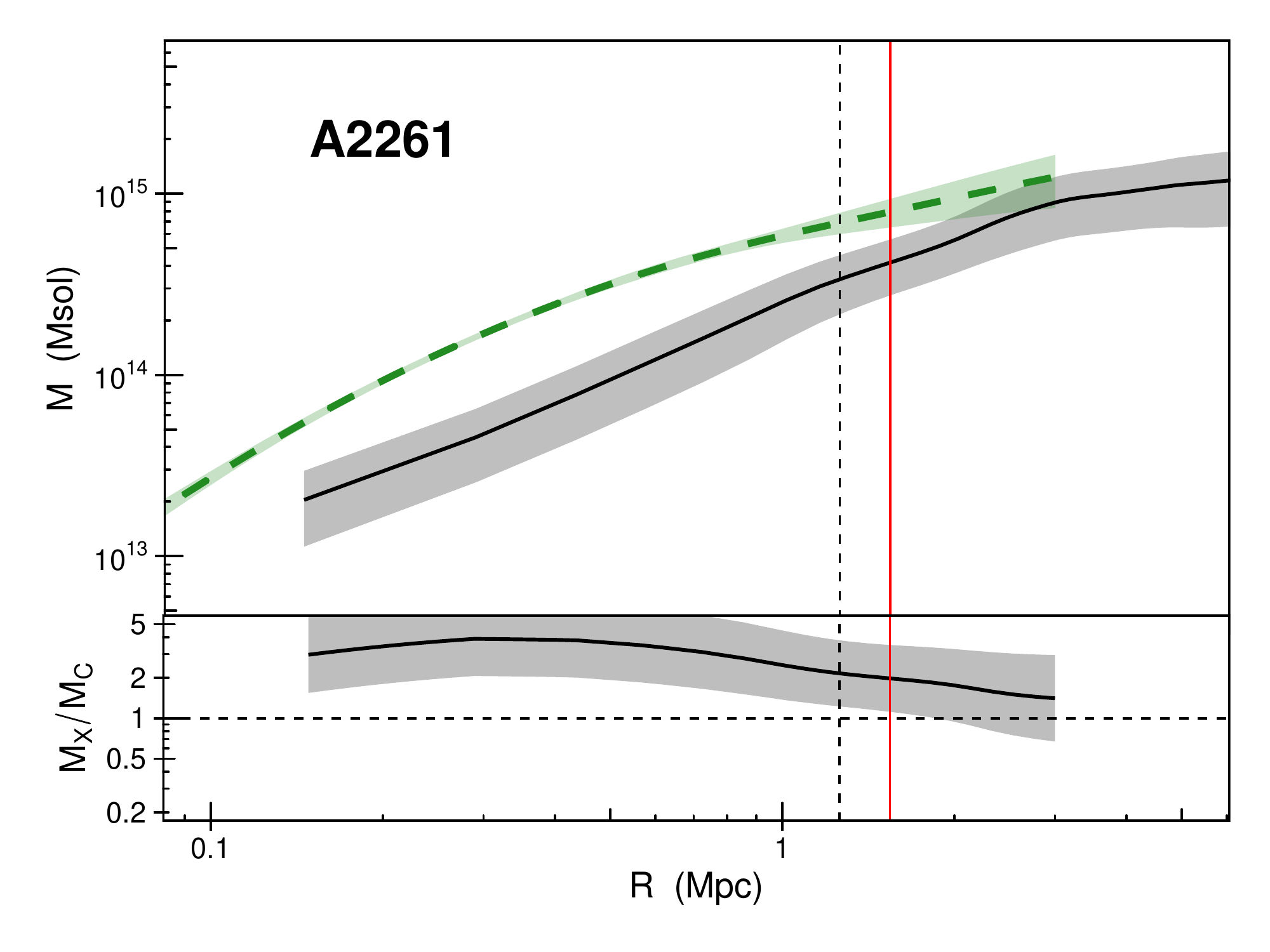} \\
\includegraphics[angle=0,width=210px]{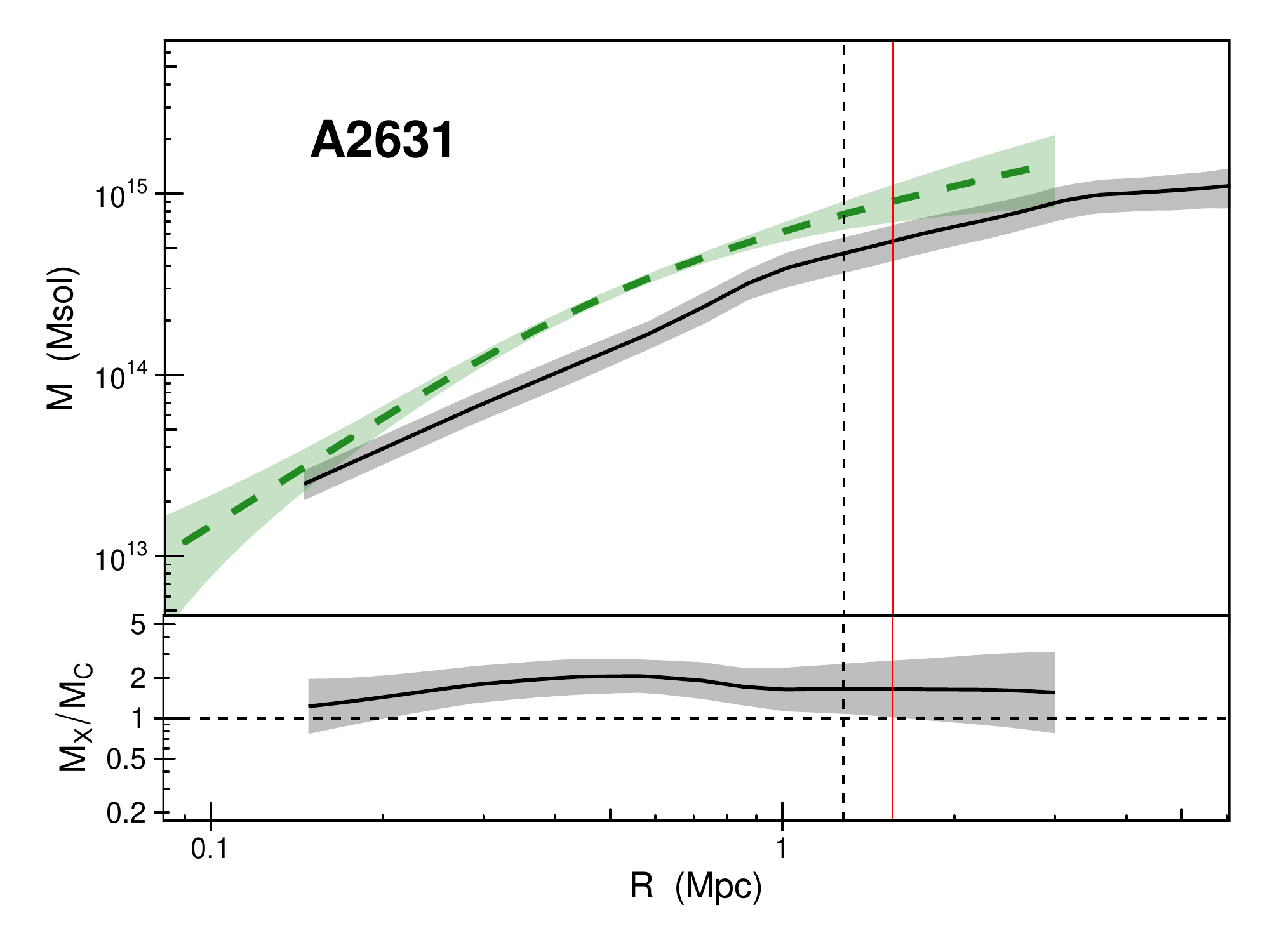} \qquad
\includegraphics[angle=0,width=210px]{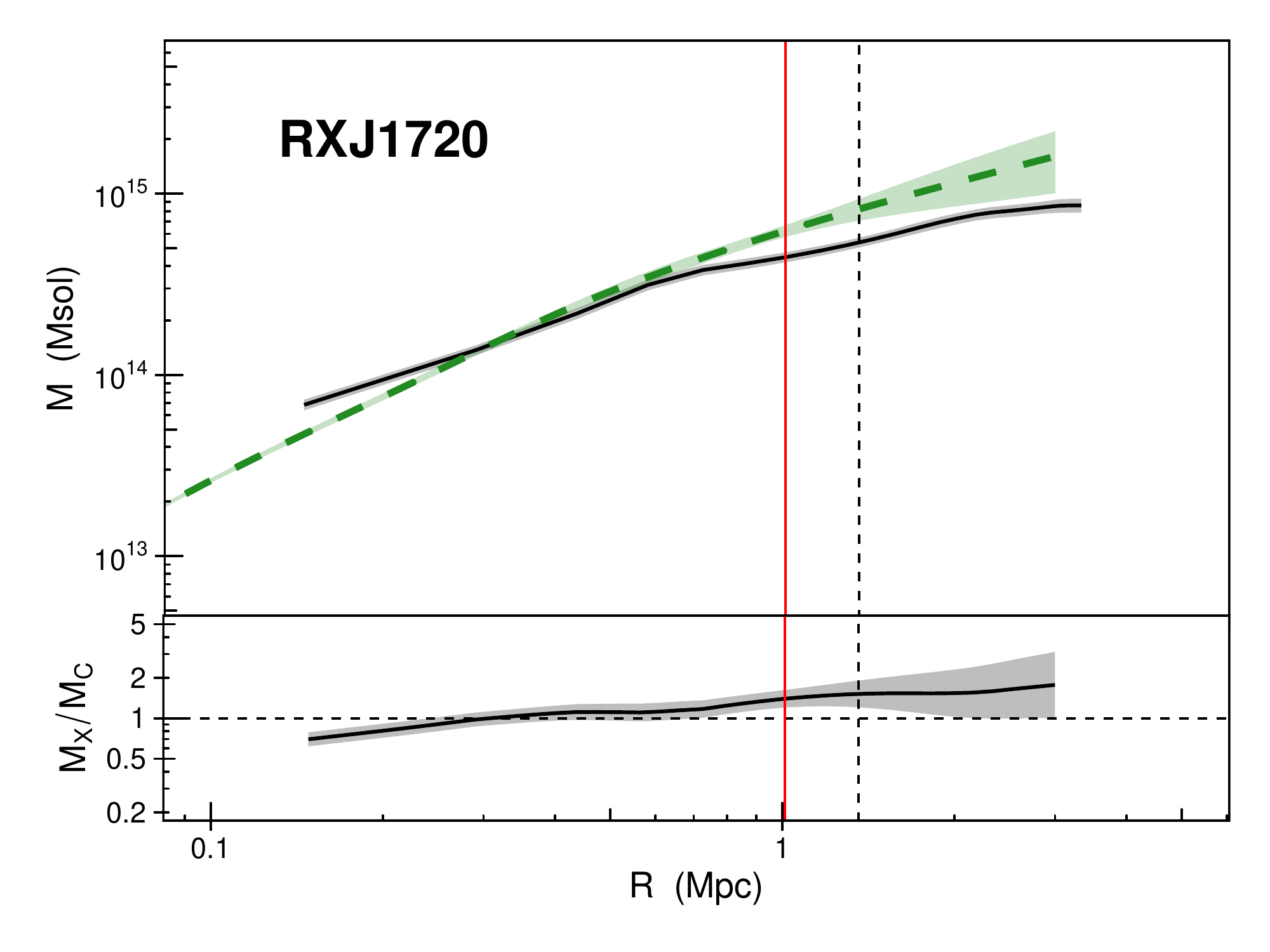} \\
\includegraphics[angle=0,width=210px]{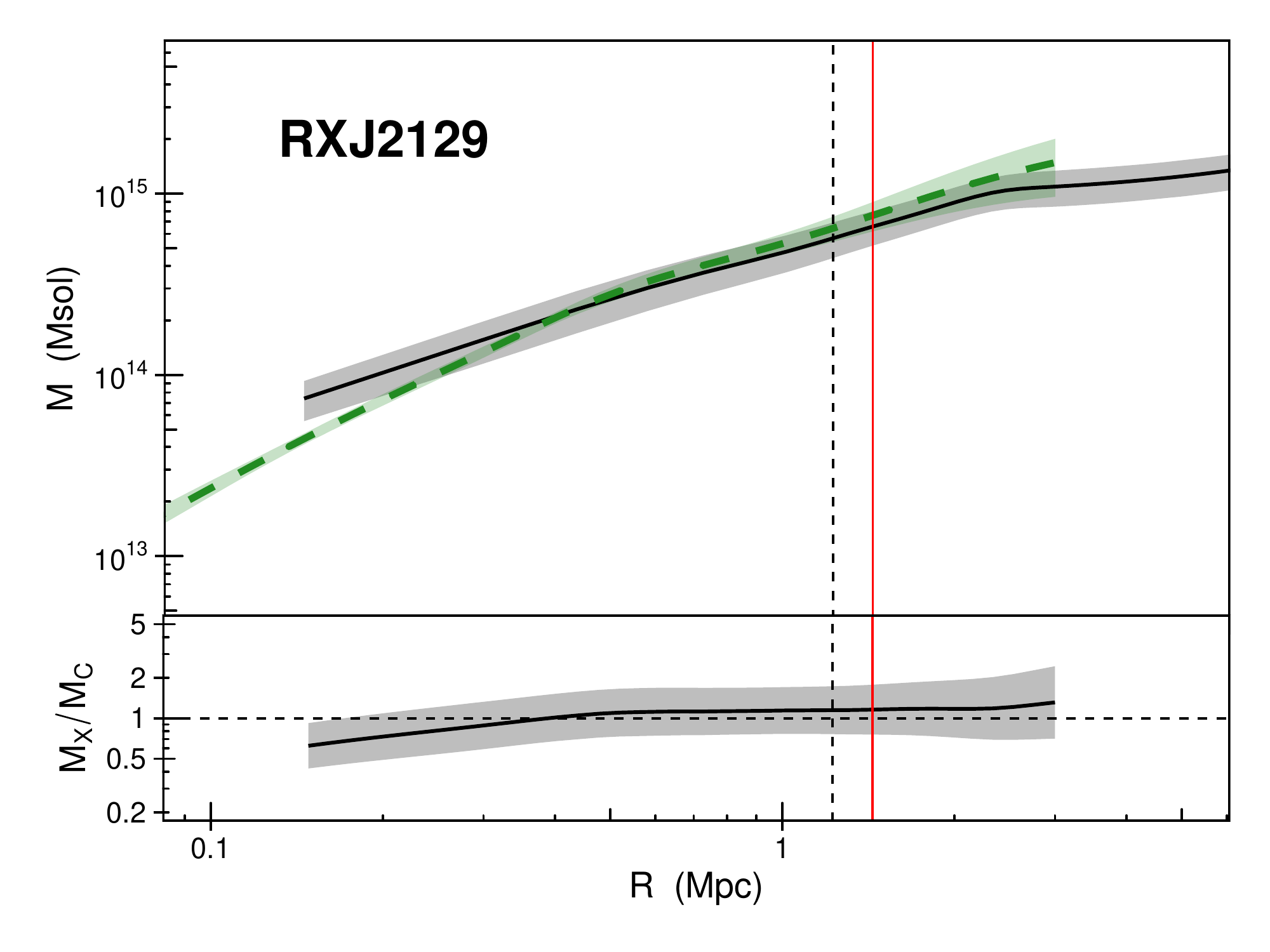} \qquad
\includegraphics[angle=0,width=210px]{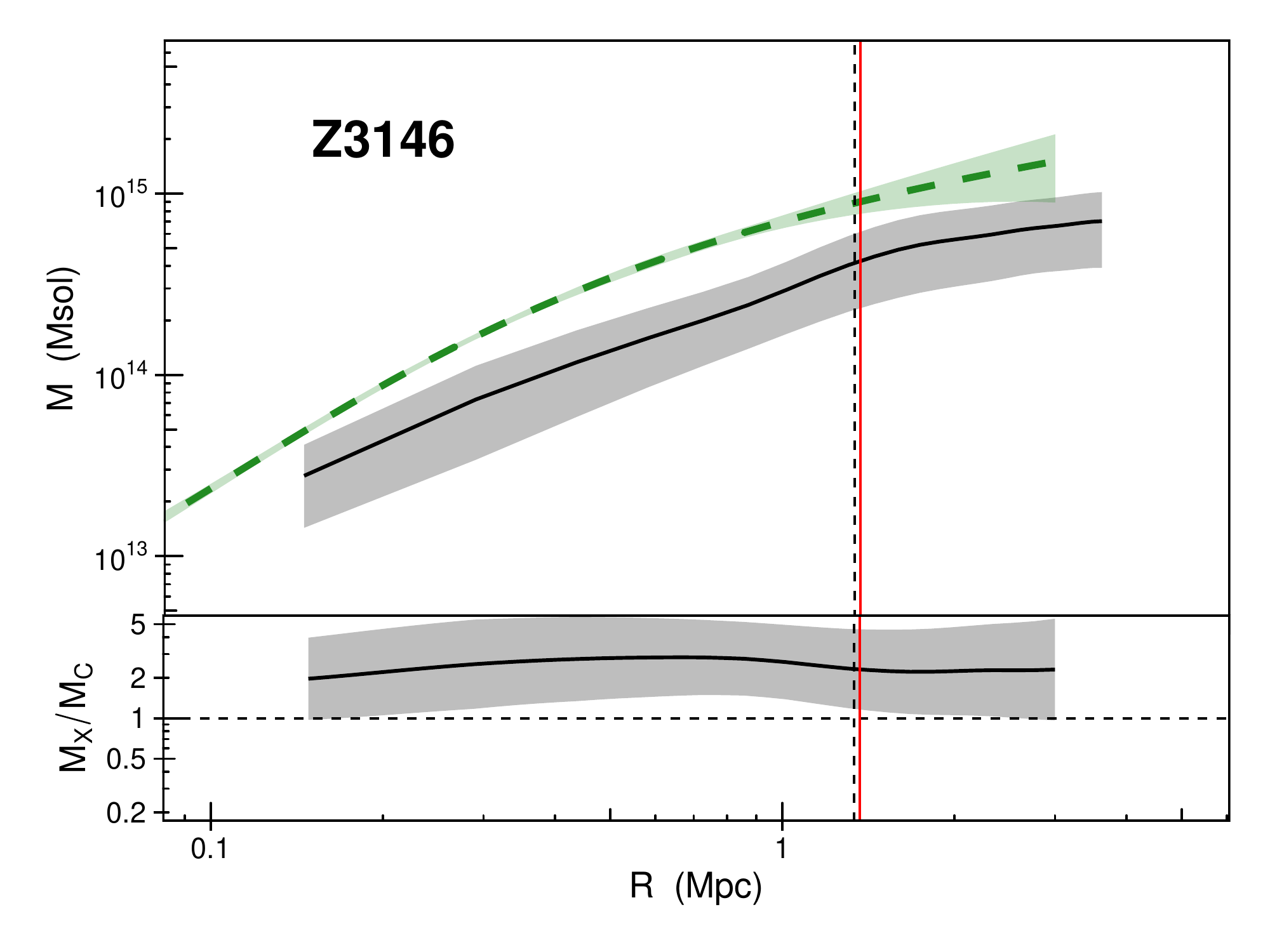}
\caption{\label{fig.xcprofs2} The upper panels in each plot show the caustic (black solid) and hydrostatic (green dashed) mass profiles for each cluster, while the lower panels show the ratio of the hydrostatic to caustic masses. The shaded regions indicate $1\sigma$ uncertainties. The vertical black dashed line indicates the value of $\rf$ estimated from the hydrostatic profile, and the vertical red solid line indicates the extent of the measured temperature profile - hydrostatic masses beyond this radius are based on extrapolation.}
\end{figure*}

Figures \ref{fig.xcprofs1} and \ref{fig.xcprofs2} show the hydrostatic
and caustic mass profiles for each cluster in our sample.

\end{document}